\documentclass[11pt]{article}

\usepackage{mathrsfs}
\usepackage[utf8]{inputenc}
\usepackage{epsfig,graphicx,color,mathtools}
\usepackage{subfig}
\usepackage{latexsym}                  
\usepackage{amsmath,amsfonts,verbatim,xkeyval,bm, amsthm, upgreek, amscd, wasysym, amssymb}
\usepackage{cases}
\usepackage{cancel}

\usepackage[english]{babel}
\usepackage{listings}
\usepackage{multicol}
\usepackage{tikz-cd}
\usepackage{booktabs}
\usepackage[numbers, sort&compress]{natbib}
\usepackage{hyperref} 
\hypersetup{hidelinks} 
\usepackage{sidecap}
\usepackage[font=small,labelfont=bf]{caption}
\usepackage{float}
\usepackage{titlesec}
\usepackage{soul}
\usepackage[normalem]{ulem}
\newcommand{\stkout}[1]{\ifmmode\text{\sout{\ensuremath{#1}}}\else\sout{#1}\fi}

\titleformat{\paragraph}
{\normalfont\normalsize\bfseries}{\theparagraph}{1em}{}
\titlespacing*{\paragraph}
{0pt}{3.25ex plus 1ex minus .2ex}{1.5ex plus .2ex}
 
\def\vet#1{{\bm #1}}

\def\Ascr{\mathcal{A}}

\def\Gscr{\mathcal{G}}
\def\Hscr{\mathcal{H}}

\def\Lscr{\mathcal{L}}

\def\Oscr{\mathcal{O}}
\def\Pscr{\mathcal{P}}
\def\Rscr{\mathcal{R}}

\def\Zscr{\mathcal{Z}}

\def\rho{\varrho}

\def\poisson#1#2{\lbrace #1,#2 \rbrace}

\def\cgrav{\Gscr}

\def\ua{\underline{a}}
\def\ue{\underline{e}}
\def\ui{\underline{i}}
\def\uell{\underline{\ell}}
\def\ug{\underline{g}}
\def\uh{\underline{h}}
\def\uL{\underline{L}}
\def\uG{\underline{G}}
\def\uH{\underline{H}}
\newcommand{\Moon}{\leftmoon}
\newcommand{\Earth}{\oplus}
\newcommand{\Sun}{\astrosun}

\title{Fully analytical propagator for lunar satellite orbits in closed form}
\author{
  {\bf Rita Mastroianni}\\
  {\small Advanced Concepts Team, European Space Agency, European Space Technology and Research Centre}\\
  {\small Keplerlaan 1, 2201 AZ Noordwijk}\\
  {\bf Edoardo Legnaro}\\
  {\small MIDA, Dipartimento di Matematica, Universit\`a di Genova}\\
  {\small via Dodecaneso 35, 16146 Genova,}\\
  {\bf Christos Efthymiopoulos}\\
  {\small Dipartimento di Matematica ``Tullio Levi-Civita'', Universit\`a degli Studi di Padova,}\\
  {\small via Trieste 63, 35121 Padova,}\\
  [1.2ex]
  {\small e-mails:
  {\tt rita.mastroianni@esa.int, edoardo.legnaro@edu.unige.it, cefthym@math.unipd.it}}
}
\date{}

\begin{document}

\maketitle

\noindent
{\small\bf Abstract:} {\small We present a fully analytical propagator for the orbits of lunar artificial satellites in a lunar gravity and third-body model sufficiently precise for a wide range of practical applications. The gravity model includes the twelve most important lunar gravity harmonics as well as the Earth's quadrupole tidal terms with a precise representation of the Earth's lunicentric ephemeris, and it gives an accuracy comparable to the way more extended semi-analytical propagator SELENA \cite{eftetal2023} for satellite orbits at altitudes from $300$ to $\sim 3000$~km. Extra terms of a more complete gravity model are straightforward to include using the formulas of the presently discussed analytical theory. The theory is based on deriving an approximate analytical solution of the secular part of the equations of motion using a Hamiltonian normal form in closed form. In total, we have two types of element transformations: from osculating to mean elements (as in \cite{eftetal2023}), and from mean to proper elements. The solution of the problem in proper elements is trivial, and, through the inverses of the above transformations, it allows to recover the position and velocity of a satellite analytically at any time $t$ given initial conditions of the osculating elements at time $t_0$ without any intermediate numerical propagation. The propagator model is valid in time spans of several decades, and for every initial condition leading to no-fall on the Moon's surface, except for identified thin zones around a set of secular resonances corresponding to commensurabilities between the satellite's secular frequencies and the secular frequencies of the lunicentric Earth's orbit. Open software python and symbolic routines implementing our propagator are provided in the repository~\cite{legmaseft2025}. Precision tests with respect to fully numerical orbital propagation in Cartesian coordinates are reported.}

\section{Introduction}
\label{sec:Intro}
In the present paper we discuss the theory and computer implementation of a fully analytical model of orbit propagation for lunar artificial satellites in a precise gravity and third-body model which is a refinement of the model described in~\cite{legeft2024}. 

The theory developed in the following sections leads to direct formulas allowing for the analytical propagation of a satellite's mean orbital elements, i.e., the elements which appear in the equations of motion after averaging with respect to short-period terms. All formulas presented below are produced in `closed form', i.e., without any expansion of the equations of motion in the orbital eccentricity. To this end, use is made of  the closed-form algebra for the computation of Poisson brackets presented in \cite{eftetal2023}. Collecting together the formulas of the whole procedure, i.e.:

(i) the closed form transformations from osculating to mean elements and their inverses, 

(ii) the closed-form transformations from mean to \textit{proper elements} (i.e. the elements of the analytical secular theory) and their inverses, and 

(iii) the analytical propagation formulas for the proper elements, 

\noindent 
leads to a theory of propagation of lunar satellite orbits based only on analytical formulas. This is accompanied with a complete package of computer routines implementing the theory, which is contained in the repository~\cite{legmaseft2025}. These routines are designed to serve a user's development of own orbital propagator. While our package refers only to the terms included in our present gravity model, producing additional terms with a more complete gravity model is straightforward using the formulas of the present paper for the procedure steps (ii) and (iii) above, as well as the formulas provided in \cite{eftetal2023} for the step (i) above.

The approximations used in the proposed gravity model imply some limits of applicability, beyond which the precision of our present formulas is not guaranteed. As regards the satellite's semi-major axis, our theory is shown below to be in general precise beyond $a_{min}\approx R_{\Moon}+300~$km and up to $a_{max}\approx R_{\Moon}+3000~$km. The lowermost limit stems from the effect of lunar mascons, which, for altitudes $a<a_{min}$, renders necessary to consider many more harmonics of the lunar gravity model than those presently used. However, we still obtain precise orbits at low altitudes for such initial conditions (e.g. polar orbits) for which the eccentricity does not undergo large variations. The highermost limit stems, instead, mostly from the exclusion from the model of the P3 and P4 terms of the Earth's tide, and also, to some extent, by a resonance overlap effect among particular secular resonances of the problem, explained in detail below. As regards the limits of applicability with respect to initial conditions in the inclination-eccentricity $(i,e)$ plane, for fixed semi-major axis $a$, these come from two sources. On one hand, initial conditions leading to collision with the Moon's surface must be excluded. The permissible (no-fall) domain in the $(i,e)$ plane  as a function of $a$ can be computed analytically by the formulas given in \cite{legeft2024}. On the other hand, our presently implemented algorithm for computing the secular normal form is `non-resonant', implying that our theory leads to very small divisors within thin zones around the few identified secular resonances which are found to cross the $(i,e)$ plane at fixed values of $a$ within the above defined range $a_{min}\leq a \leq a_{max}$. Such resonances stem from commensurabilities between three distinct types of frequencies: i) the satellite's precession frequencies $\dot{g},\dot{h}$, ii) the Moon's rotation frequency $\omega_{\Moon,z}$, and iii) the secular frequencies associated with the Earth's lunicentric orbit. The latter are the solar year frequency $\dot{\ell}_{\Sun}$, and the frequencies of precession of the Moon's perigee $\dot{\varpi}_{\Moon}$ and line of nodes in the ecliptic $\dot{h}_{\Moon}$. A detailed account of the effect of these commensurabilities is given in section \ref{subsec:resonances} below. 

Besides the usual elimination of the `fast' degree of freedom, associated with a satellite's mean anomaly $\ell$, analytical theories of lunar satellite orbits encounter the particularity of the presence of medium-period terms in the equations of motion associated with the Moon's slow rotation. In a lunicentric body frame such as PALRF (see references in \cite{eftetal2023}), the medium period angle is the satellite's longitude of the nodes $h$. Short-period terms independent of $h$ arise from the Moon's zonal harmonics, and can be easily normalized in closed form using the techniques discussed in~\cite{bro1959}, ~\cite{desaehen2004a} and~\cite{desae2006b}, ~\cite{laretal2009a}, ~\cite{laretal2009b}, ~\cite{laretal2020}, and thoroughly implemented in \cite{eftetal2023}. On the other hand, terms with combinations of the short and medium period angles $\ell,h$ arise due both to the lunar potential's tesseral harmonics and the Earth's tide as computed in the Moon's body frame. Such terms can be eliminated essentially using relegation techniques (see \cite{desae2006b} and \cite{eftetal2023}). The same techniques are applicable to the elimination of solar tide and solar radiation pressure terms, whose importance, however, is anyway limited for lunar satellites. Alternative to the usual closed form and relegation techniques is the approach discussed in~\cite{mahalf2019} and ~\cite{mah2024}. While still referred to by the authors as a `closed-form' approach, one can notice that the closed form computations in the latter references require numerical computation of several integrals (`quadratures'), a process which in practice prevents the method from being  completely analytical. In the context of our own theory, instead, the term `fully analytical' means that the evaluations performed at any step of the theory should involve only algebraic operations between the terms of series of the polynomial-trigonometric form. 

As discussed already in~\cite{giaetal1970}, after eliminating the short and medium period terms depending on $\ell$ and $h$ respectively, the final `secular' Hamiltonian now depends only on the satellite's orbital angle $g$ (argument of the perilune). The resulting equations of motion can then be regarded as a one degree of freedom (DOF) system which is in principle integrable. Hence, the equations of motion can be either integrated directly or by using some other kind of analytical approximation based on quadratures (for instance, by elliptic integrals). On the other hand, a completely analytical theory can arise after eliminating also the angle $g$ using some canonical normalization procedure. Examples of such fully analytical theories in the literature are: i) the theory of ~\cite{knemil1998}, which deals with low (altitude smaller than $100$ [Km]) polar lunar orbits. This theory takes into account only the lunar gravity field, removing first both the short and medium periodic perturbations and then analytically integrating the resulting dynamical system expanded up to just the second order in the orbital eccentricity. ii) In~\cite{laretal2009a} a `closed-form' analytical theory is derived using a $50\times 0$ gravity field and including third-body perturbation, which is valid for almost-planar orbits, generalized in~\cite{laretal2009b} to orbits at any inclination. The elimination of the short and medium period angles $\ell$ and $h$ in this model leads to a Hamiltonian depending only on the angle $g$. However, a general analytical solution of the resulting secular equations of motion involves huge expressions based on elliptic and hyper-elliptic integrals. Due to this, the authors apply their theory essentially only to some orbits of particular interest in astrodynamics, namely the frozen orbits which are equilibria of the secular equations of motion. iii) The most accurate, to our knowledge, theories are developed in~\cite{mahalf2019} and ~\cite{mah2024}. The theory discussed in~\cite{mahalf2019} refers only to the lunar gravitational terms and consists of three separate transformations: first, the short-period terms due to zonals and tesserals are separated from the Hamiltonian using two transformations, and then, the long-period terms are separated using the long-period transformation to obtain a completely secular Hamiltonian. The theory was recently generalized in~\cite{mah2024} to include lunar gravity spherical harmonics up to an arbitrary degree and order as well as the third-body effects of the Earth and Sun. Here as well three separate canonical transformations are performed, sequentially eliminating short-period, medium-period, and long-period terms. As commented above, in the first transformation, the averaging of tesseral short-period terms involves a solution of the homological equation which, despite being called `closed-form', actually is expressed in terms of two quadratures which are to be computed numerically. Furthermore, third-body terms are handled through the usual elliptic expansion using Hansen’s coefficients. Finally, the orbital elements of the Earth and Sun in the selenocentric equatorial frame are  treated as constants. 

Most of the limitations of the existing analytical theories discussed above are addressed by our own approach presented in the sections to follow. Section~\ref{sec:Ham_mod} presents the Hamiltonian model which serves as point of departure of our analytical theory. This includes the selection of a particular set of zonal and tesseral terms of the lunar gravity model (same as in \cite{legeft2024}) as well as the quadrupolar (P2) terms of the the Earth's tidal potential. However, contrary to~\cite{legeft2024}, as regards the latter terms here we use a more accurate representation of the Earth's lunicentric position vector, which includes quasi-periodic variations with all the basic secular frequencies important for the problem, namely the solar year frequency and the frequencies of precession of the Moon's perigee and line of nodes in the ecliptic plane. Section~\ref{sec:analytical} contains our secular theory leading to a fully analytical propagator. In particular, we present the normalization procedure used to pass from mean to proper elements, and hence to remove the dependence of the equations of motion on the secular angles of the problem. Section~\ref{sec:propagator} describes the algebraic and algorithmic structure of the propagator, referring to the use of non-singular (equinoctial) secular variables in all practical computations. Then, an error analysis of the normal form method is made. We study the error in the analytical versus numerical propagation first of the mean elements, and then of the osculating elements of a trajectory, the latter computed through an orbital integrator in cartesian coordinates. We analyze the various sources of error, playing particular emphasis on the role of secular resonances, which affect the accuracy of the analytical solutions through their appearance in small divisors of the analytical (series) terms.  Section~\ref{sec:conclusion} contains a summary of the main results from the present study. 

\section{Reference frame and equations of motion}
\label{sec:Ham_mod}
As in~\cite{eftetal2023} and \cite{legeft2024}, we adopt as reference frame the Principal Axis Lunar Reference Frame (PALRF). This is a body frame with origin at the barycenter of the Moon, z-axis equal to the Moon’s mean axis of revolution, and x-axis pointing towards the Moon’s meridian with largest equatorial radius. The equations of motion of a point mass (satellite) in the above frame are obtained as Hamilton's equations under the Hamiltonian
\begin{equation}\label{ham}
 \Hscr = \frac{1}{ 2}\mathbf{p}^2 
-\vet{\omega}_{\Moon}(t)\cdot(\mathbf{r}\times\mathbf{p}) + V(\mathbf{r},t)~~.
\end{equation}
In Eq.~\eqref{ham}, $\mathbf{r}(t)$ is the lunicentric PALRF radius vector of the satellite, $\mathbf{p}(t)$ is the velocity vector of the satellite in a rest frame whose axes coincide with the axes of the PALRF at the time $t$, $\boldsymbol\omega_{\Moon}(t)$ is the Moon's angular velocity vector, and $V(\mathbf{r},t)$ is the potential corresponding to the adopted force model. 

\subsection{Potential}
\label{subsec:potential}
The potential $V(\mathbf{r},t)$ can be written as:
\begin{equation}\label{pot}
V(\mathbf{r},t) = V_{\Moon}(\mathbf{r}) + V_{\Earth}(\mathbf{r},t)+ V_{\Sun}(\mathbf{r},t)
+V_{SRP}(\mathbf{r},t)
\end{equation}
where $V_{\Moon}$, $V_{\Earth}$, $V_{\Sun}$ and $V_{SRP}$ are, respectively, the potential terms due to the Moon's gravity, Earth and Sun tides and the solar radiation pressure. In \cite{legeft2024} several precision tests are made on the accuracy of various truncations of the above model, necessary to simplify the analytical treatment of the equations of motion. From Figure $7$ of~\cite{eftetal2023} and Figure 1 of~\cite{legeft2024}, it is evident that the Sun’s gravity and SRP becomes important only at very high altitudes and they can be ignored completely for any satellite orbit with $a<5000~$km. Furthermore, as regards the influence of the terms $V_{\Moon}(\mathbf{r})$, $ V_{\Earth}(\mathbf{r},t)$ on the dynamics, the authors of~\cite{legeft2024} distinguish four zones in altitude, corresponding to different dynamical regimes:\\
\\
\noindent
{\it Zone of essentially non-secular dynamics} (altitude $a-R_{\Moon}<100$ km): in this zone, several harmonics of the lunar potential compete in size, as a consequence of the prevalence of lunar mascons. As discussed in ~\cite{laretal2020}, zonal terms up to $n = 30$ still influence the orbits, leading to substantial short-period (and possibly chaotic) effects not consistent with the definition of `secular' orbital behavior.\\
\noindent
{\it Low Altitude Zone} of essentially secular dynamics (altitude $100$ km$\leq a-R_{\Moon}< 500$ km): in this zone we have an accumulation of the contribution of many harmonics of degree $n\leq 10$, a fact leading to a total force perturbation of about two orders of magnitude larger than the one due to the Earth's tide.\\
\noindent
{\it Middle Altitude Zone} of essentially secular dynamics (altitude $500$ km$\leq a-R_{\Moon}\leq 5000$ km): near the lower limit of this zone the lunar multipoles produce a force perturbation superior by more than one order of magnitude to the Earth's tidal perturbation. However, this is reversed at the higher limit, where the problem becomes essentially one perturbed by the Earth plus few harmonics $n=2$ and $n=3$.\\
\noindent
{\it High Altitude Zone} of essentially non-secular dynamics (altitude $a-R_{\Moon}> 5000$ km): any perturbation other than the Earth's tidal becomes negligible. The Earth’s tidal acceleration, as well as centrifugal and Coriolis accelerations dominate the dynamics beyond the altitude $\sim 2100$ km.\\
\\
Based on the above considerations, \cite{legeft2024} have proposed the following truncation of the potential $V(\mathbf{r},t)$, called the `simplified for secular-dynamics model' (SSM) as sufficiently accurate for all practical purposes in the altitudes $100$ km$\leq r-R_{\Moon}\leq 3000 $ km: $V_{\Sun}$ and $V_{SRP}$ are completely neglected. From the lunar potential 
\begin{equation}\label{potmoon}
V_{\Moon}(\mathbf{r}) = -{\mathcal{G}M_{\Moon}\over r}\sum_{n=0}^\infty
\left({R_{\Moon}\over r}\right)^n\sum_{m=0}^n P_{nm}(\sin\phi)
[C_{nm}\cos(m\lambda)+S_{nm}\sin(m\lambda)]
\end{equation}
($\mathcal{G}=$~Newton's gravity constant, $M_{\Moon}=$~Moon's mass, $R_{\Moon}=$~equatorial Moon radius, $\phi,\lambda=$~satellite's longitude and latitude (see figure 1 of~\cite{eftetal2023}), $r=$lunicentric satellite's distance, $P_{nm}=$~the normalized Legendre polynomial of degree $n$ and order $m$, $C_{nm}$ and $S_{nm} = $ zonal ($m=0$) and tesseral ($m\neq 0$) coefficients), the following set of 
twelve harmonics is included: 
\begin{equation}\label{ssmset}
    \mathcal{CS}_{SSM}=\left\{C_{20}, C_{22}, C_{30}, C_{31}, S_{31}, 
    C_{40}, C_{41}, C_{60},C_{70}, C_{71}, C_{80}, C_{90}\right\}~.
\end{equation}
Hence, the formula 
\begin{equation}\label{potmoon_simply}
V_{\Moon}(\mathbf{r})\simeq 
V^{SSM}_{\Moon}(\mathbf{r}) = -{\mathcal{G}M_{\Moon}\over r}\sum_{n=1}^9
\left({R_{\Moon}\over r}\right)^n\sum_{m=0}^n P_{nm}(\sin\phi)
[C_{nm}\cos(m\lambda)+S_{nm}\sin(m\lambda)]
\end{equation}
is adopted, with $\lbrace C_{nm},\,S_{nm}\rbrace\in \mathcal{CS}_{SSM}$.

The Earth's tidal potential is approximated by the quadrupolar term: 
\begin{equation}\label{potearth}
V_\Earth(\mathbf{r},t) \simeq V_{\Earth}^{SSM}(\mathbf{r},t) = 
V_{\Earth,P2}\left(\mathbf{r},\mathbf{r}_{\Earth}(t)\right)=
{\mathcal{G}M_{\Earth} \over r_{\Earth}(t)}
\left(
{1\over 2}{r^2\over r_{\Earth}^2(t)}
-{3\over 2}{(\mathbf{r}\cdot\mathbf{r}_{\Earth}(t))^2\over r_{\Earth}^4(t)}
\right)\,\, ,
\end{equation}
where the lunicentric PALRF radius vector of the Earth $\mathbf{r}_{\Earth}(t)$ is computed by the Fourier representation of the  high-accuracy ephemerides (INPOP19a) reported in~\cite{eftetal2023}. In~\cite{legeft2024}, a simple analytical model (see their equations (19)) was adopted, which includes only the first two Fourier terms of the above representation and is physically equivalent to the Earth's lunicentric orbit being a fixed ellipse of semi-major axis $a_{\Earth}=382470~$km, eccentricity $e=0.0547$ and mean motion equal to the Moon's rotation frequency, so that the guiding center of the Earth's epicycle is along the $x-$axis of the PALRF frame. Here, instead, we adopt a still simple, but more precise model which has a maximum error of $\sim 10^3~$km on the Earth's PALRF position vector $\mathbf{r}_{\Earth}$ with respect to the high-accuracy ephemeris. First, we set
\begin{equation}\label{rbearth}
\mathbf{r}_{\Earth}(t) = (x_{\Earth}(t),y_{\Earth}(t),z_{\Earth}(t))
\end{equation}
with 
\begin{equation}
\label{rE_approx}
\begin{aligned}
&x_{\Earth}(t)\approx 
{A}^{(x)}_1 +\sum_{i\in\left\lbrace 15,\, 20,\, 32,\, 37\right\rbrace  }\left({A}^{(x)}_i \cos\omega^{(x)}_i t + {B}^{(x)}_i \sin\omega^{(x)}_i t\right)\, ,\\
& y_{\Earth}(t)\approx 
\sum_{i\in\left\lbrace 8,\, 19,\, 23,\, 34,\, 39\right\rbrace  }\left({A}^{(y)}_i \cos\omega^{(y)}_i t + {B}^{(y)}_i \sin\omega^{(y)}_i t\right)\, ,\\
& z_{\Earth}(t)\approx 
\sum_{i\in\left\lbrace 5,\, 15,\, 23\right\rbrace  }\left({A}^{(z)}_i \cos\omega^{(z)}_i t + {B}^{(z)}_i \sin\omega^{(z)}_i t\right)\, ,
\end{aligned}
\end{equation}
and constants equal to those reported in Tables $1$, $2$, $3$ of~\cite{eftetal2023}, namely
\begin{equation} 
\label{x_earth}
\begin{aligned}
 \lbrace A_{1}^{(x)}, \, A_{15}^{(x)},\, A_{20}^{(x)},\, A_{32}^{(x)},\, A_{37}^{(x)}  \rbrace &= \lbrace 382469.63,\,730.56,\,14796.88,\, 1369.26\, ,-1285.72 \rbrace  \, [\mathrm{km}]\, ,\\
 \lbrace  B_{15}^{(x)},\, B_{20}^{(x)},\, B_{32}^{(x)},\, B_{37}^{(x)}  \rbrace &= \lbrace 3836.11,\, 14794.17,\,-2010.23,\,148.03 \rbrace  \, [\mathrm{km}]\, ,\\
 \lbrace  \omega_{15}^{(x)},\, \omega_{20}^{(x)},\, \omega_{32}^{(x)},\, \omega_{37}^{(x)}  \rbrace &= \lbrace 0.19751,\,0.228027,\,0.425537,\,0.461792 \rbrace  \, [\mathrm{rad/days}]\, ,
\end{aligned}
\end{equation}
\begin{equation} 
\label{y_earth}
\begin{aligned}
 \lbrace  \, A_{8}^{(y)},\, A_{19}^{(y)},\, A_{23}^{(y)},\, A_{34}^{(y)},\, A_{39}^{(y)}  \rbrace &=\lbrace 60.59,\, 8406.78,\,29773.72,\, -3262.64,\, 148.26  \rbrace  \, [\mathrm{km}]\,, \\
 \lbrace  B_{8}^{(y)},\, B_{19}^{(y)},\, B_{23}^{(y)},\, B_{34}^{(y)},\, B_{39}^{(y)}  \rbrace &=\lbrace -1403.61,\, -1596.08,\, -29749.79,\,-2223.9,\,1287.76    \rbrace  \, [\mathrm{km}]\,, \\
 \lbrace  \omega_{8}^{(y)},\, \omega_{19}^{(y)},\, \omega_{23}^{(y)},\, \omega_{34}^{(y)},\, \omega_{39}^{(y)}  \rbrace &= \lbrace 0.017202,\, 0.19751,\, 0.228027,\, 0.425537,\, 0.461792 \rbrace  \, [\mathrm{rad/days}] \,,
\end{aligned}
\end{equation}
\begin{equation} 
\label{z_earth}
\begin{aligned}
 \lbrace   A_{5}^{(z)},\, A_{15}^{(z)},\, A_{23}^{(z)}  \rbrace &=  \lbrace  -2579.36,\, -825.59,\,  -44649.42 \rbrace  \, [\mathrm{km}]\,, \\
 \lbrace  B_{5}^{(z)},\, B_{15}^{(z)},\, B_{23}^{(z)}  \rbrace &=  \lbrace 2895.76,\, 1073.41,\, 2554.55 \rbrace  \, [\mathrm{km}]\,, \\
\lbrace  \omega_{5}^{(z)},\, \omega_{15}^{(z)},\, \omega_{23}^{(z)}  \rbrace &= \lbrace 0.00286858, \,0.194642,\, 0.230895  \rbrace  \, [\mathrm{rad/days}] \, .
\end{aligned}
\end{equation}
In Eqs.(\ref{rE_approx}), the time $t=0$ corresponds to 12 noon of the Julian Day 2000. Furthermore, we readily check that all the frequencies appearing in the above tables can be derived as linear combinations of four \textit{fundamental frequencies}. We have:
\begin{equation}
\label{lin_comb_fund_freq}
\begin{aligned}
&\omega_{15}^{(x)}=\omega_{19}^{(y)}\simeq \dot{\varphi}_1 +\dot{\varphi}_2-2\dot{\varphi}_4 \, , &\qquad\quad
&\omega_{20}^{(x)}=\omega_{23}^{(y)}\simeq \dot{\varphi}_1 -\dot{\varphi}_2 \, ,\\
&\omega_{32}^{(x)}=\omega_{34}^{(y)}\simeq 2\left(\dot{\varphi}_1 -\dot{\varphi}_4\right) \, , &\qquad\quad
&\omega_{37}^{(x)}=\omega_{39}^{(y)}\simeq 2\left(\dot{\varphi}_1 -\dot{\varphi}_3\right) \, ,\\
&\omega_{8}^{(y)}\simeq \dot{\varphi}_4  \, ,&\qquad\quad
&\omega_{5}^{(z)}\simeq \dot{\varphi}_2 -\dot{\varphi}_3 \, ,\\
&\omega_{15}^{(z)}\simeq \dot{\varphi}_1 +\dot{\varphi}_3 - 2 \dot{\varphi}_4 \, ,&\qquad\quad
&\omega_{23}^{(z)}\simeq \dot{\varphi}_1 -\dot{\varphi}_3 \, ,
\end{aligned}
\end{equation}
where 
\begin{equation}
\label{fund_freq}
\begin{aligned}
&\dot{\varphi}_1=\dot{\lambda}_{\Moon}=2\pi/(27.322)=0.229968\,\, [\mathrm{rad/days}]\,  \\
&\dot{\varphi}_2=\dot{\varpi}_{\Moon}=2\pi/(8.84\cdot 365.26)=0.0019443\,\, [\mathrm{rad/days}]\,\\
&\dot{\varphi}_3=\dot{h}_{\Moon}=-2\pi/(18.6\cdot 365.26)=-0.000924193\,\, [\mathrm{rad/days}]\,\\
&\dot{\varphi}_4=\dot{\ell}_{\Sun}=2\pi/(365.26)=0.017202\,\, [\mathrm{rad/days}]\, .
\end{aligned}
\end{equation}
Computing the periods associated with the four fundamental frequencies, we recognize that $2\pi/\dot{\lambda}_{\Moon}\approx 27.322$~days is the siderial lunar month, $2\pi/\dot{\varpi}_{\Moon}\approx 8.84$~years is the precession of the \textit{longitude} of pericenter $\varpi$ of the Moon's geocentric orbit in the ecliptic frame (sometimes erroneously reported as the period of precession of the argument of the pericenter), $-2\pi/\dot{h}_{\Moon}\approx 18.6$ years is the frequency of precession of the longitude of the nodes of the Moon's geocentric orbit in the ecliptic frame and $2\pi/\dot{\ell}_{\Sun}\approx 365.25$ days is the Solar year. Consider, finally, the angles: 
\begin{align}\label{phases}
\varphi_1 &= \varphi_{1,0}+\dot{\lambda}_{\Moon}t\, , \nonumber \\
\varphi_2 &= \varphi_{2,0}+\dot{\varpi}_{\Moon}t\, ,  \\
\varphi_3 &= \varphi_{3,0}+\dot{h}_{\Moon}t\, , \nonumber \\
\varphi_4 &= \varphi_{4,0}+\dot{\ell}_{\Sun}t~~\, . \nonumber 
\end{align}
Using the ratios $A_i/B_i$ of the various coefficients reported in Eqs.~\eqref{x_earth},\eqref{y_earth} and \eqref{z_earth}, we can compute the corresponding initial phases $\varphi_{i,0}$, $i=1,2,3,4$ of the angles $\varphi_i$ at 12 noon of the JD2000. Computing also the Fourier amplitudes $C_i=(A_i^2+B_i^2)^{1/2}$ we are led to the following compact final model of the Earth's PALRF radius vector, equivalent to the one reported in Eq.~\eqref{rE_approx}:
\begin{equation}\label{rE_new}
\begin{aligned}
x_{\Earth}(t) &= x_{\Earth}(\varphi_1(t),\varphi_2(t),\varphi_3(t),\varphi_4(t)) \\
&= 
C_1^{(x)} 
- C_{15}^{(x)}\cos(\varphi_1+\varphi_2-2\varphi_4) 
+ C_{20}^{(x)}\cos(\varphi_1-\varphi_2) \\
&+ C_{32}^{(x)}\cos(2\varphi_1-2\varphi_4) 
+ C_{37}^{(x)}\cos(2\varphi_1-2\varphi_3) \, ,\\
&~\\
y_{\Earth}(t) &= y_{\Earth}(\varphi_1(t),\varphi_2(t),\varphi_3(t),\varphi_4(t)) \\
&= 
C_{8}^{(y)}\cos(\varphi_4) 
+ C_{19}^{(y)}\sin(\varphi_1+\varphi_2-2\varphi_4) \\
&- C_{23}^{(y)}\sin(\varphi_1-\varphi_2) 
- C_{34}^{(y)}\sin(2\varphi_1-2\varphi_4) 
- C_{39}^{(y)}\sin(2\varphi_1-2\varphi_3)\, ,\\
&~\\
z_{\Earth}(t) &= z_{\Earth}(\varphi_1(t),\varphi_2(t),\varphi_3(t),\varphi_4(t)) \\
&= 
- C_{5}^{(z)}\sin(\varphi_2-\varphi_3) 
+ C_{15}^{(z)}\sin(\varphi_1+\varphi_3-2\varphi_4) 
- C_{23}^{(z)}\sin(\varphi_1-\varphi_3)\, ,\\
\end{aligned}
\end{equation}
with the angles $\varphi_i$ given by Eq.~\eqref{phases} and the numerical constants
\begin{equation*}
\begin{aligned}
\lbrace
C_{1}^{(x)},C_{15}^{(x)},C_{20}^{(x)},C_{32}^{(x)},C_{37}^{(x)}
\rbrace
&=
\lbrace 
382469.63, 3905.06, 20924.03, 2432.26, 1294.21
\rbrace  \, [\mathrm{km}]\,, \\
\lbrace
C_{8}^{(y)},C_{19}^{(y)},C_{23}^{(y)},C_{34}^{(y)},C_{39}^{(y)}
\rbrace
&=
\lbrace 
1404.92, 8556.95, 42089.48, 3948.49, 1296.27
\rbrace  \, [\mathrm{km}]\,, \\
\lbrace
C_{5}^{(z)},C_{15}^{(z)},C_{23}^{(z)}
\rbrace
&=
\lbrace 
3877.95, 1354.18, 44722.44
\rbrace  \, [\mathrm{km}]\,, \\
\lbrace
\varphi_{1,0},\varphi_{2,0},\varphi_{3,0},\varphi_{4,0}
\rbrace
&=
\lbrace 
-1.12751856,-0.34221198,-2.75562949,1.52765585
\rbrace \,\, .
\end{aligned}
\end{equation*}
Finally, the Moon's angular velocity vector $\mathbf{\omega}_{\Moon}$ projected in the PALRF frame can be approximated as 
\begin{align}
\label{ang_vel_Moon}
&\omega_{\Moon, x} \simeq 0 \, , & &\omega_{\Moon, y} \simeq 0\, , & 
\omega_{\Moon, z} \simeq \dot{\varphi}_1 = 0.229968 \,\, [\mathrm{rad/days}]~~.
\end{align}

\subsection{Secular Hamiltonian}
\label{subsec:hamiltonian}
Setting $V(\vet{r},t)=V_{\Moon}^{SSM}(\vet{r}) + V_{\Earth}^{SSM}(\vet{r}, t)$ in the Hamiltonian~\eqref{ham}, with the potentials $V_{\Moon}^{SSM}(\vet{r}), V_{\Earth}^{SSM}(\vet{r}, t)$ computed by the formulas of subsection \ref{subsec:potential}, we arrive at a Hamiltonian model in Cartesian coordinates and momenta $(x,y,z,p_x,p_y,p_z)$ whose equations of motion can be solved numerically to obtain satellite trajectories for any initial condition in the PALRF frame. 

To proceed with an analytical theory, instead, the Hamiltonian is first expressed in 
Delaunay canonical action angle variables 
\begin{equation} \label{delaunay}
L = \sqrt{\mathcal{G} M_{\Moon} a}, \quad G = L\sqrt{1-e^2}, \quad H= G \cos i, \quad \ell=M, \quad g=\omega, \quad h=\Omega,
\end{equation}
via the Keplerian elements $(a,e,i)$ (satellite orbit's semi-major axis, eccentricity and inclination), with $a=a(L)$, $e=e(L,G)$, $i=i(L, G,H)$ as well as the angles $(f,g,h)$ (where $f=f(\ell,e)=$ true anomaly, $g=$ argument of perilune and $h=$ longitude of the nodes), for the potential term $V_{\Moon}$, or $(u,g,h)$ ($u=u(\ell,e)=$~eccentric anomaly) for the potential term $V_{\Earth}$. As in \cite{eftetal2023}, we then consider a first canonical transformation from osculating to mean elements
\begin{equation}\label{tramean}
\mathcal{X}: (a,e,i,\ell,g,h)
\rightarrow 
(\ua,\ue,\ui,\uell,\ug,\uh)
\end{equation}
aiming to average the Hamiltonian with respect to the fast angle $\ell$. The transformation (\ref{tramean}) is computed in `closed form', i.e., without expansions in the orbital eccentricity, using the set of formulas provided in \cite{eftetal2023}. From the standard theory of transformations in closed form, we then have that, applying the transformation \eqref{tramean} to the Hamiltonian $\Hscr(a,e,i,\ell,g,h)$ leads to
\begin{equation}\label{hamsec}
\Hscr(a,e,i,\ell,g,h) = \Hscr_{sec}(\ua,\ue,\ui,\ug,\uh)
\end{equation} 
with
\begin{equation}\label{hamsm}
\Hscr_{sec}(\ua,\ue,\ui,\ug,\uh,\varphi_1,\varphi_2,\varphi_3,\varphi_4)
=
-{\mathcal{G} M_{\Moon}\over 2\ua}-\omega_{\Moon,z}\uH
+\underline{V}_{\Moon}^{SSM}
+\underline{V}_{\Earth}^{SSM}
\end{equation}
where
$$
\underline{V}_{\Moon}^{SSM}(\ua,\ue,\ui,\ug,\uh)=\Bigg(
{1\over 2\pi}\int_{0}^{2\pi} {r^2\over a^2\eta}V_{\Moon}^{SSM}df
\Bigg)_{a=\ua,e=\ue,i=\ui,g=\ug,h=\uh}
$$
$\eta=\sqrt{1-e^2}$ being the `eccentricity function', and 
$$
\underline{V}_{\Earth}^{SSM}(\ua,\ue,\ui,\ug,\uh,\varphi_1,\varphi_2,\varphi_3,\varphi_4)=\Bigg(
{1\over 2\pi}\int_{0}^{2\pi} (1-e\cos u)V_{\Earth}^{SSM}du
\Bigg)_{a=\ua,e=\ue,i=\ui,g=\ug,h=\uh}~~.
$$

The Hamiltonian $\Hscr_{sec}$ depends explicitly on time through the dependence of the potential term $\underline{V}_{\Earth}^{SSM}$ on the angles $\varphi_i$, $i=1,2,3,4$ which enter into the Earth's radius vector $\mathbf{r}_{\Earth}$. To fulfill the requirements of our analytical theory, we transform this Hamiltonian into another which is formally autonomous and Fourier expanded in all the angles of the problem, namely the angles $(\ug,\uh,\varphi_1,\varphi_2,\varphi_3,\varphi_4)$. To this end, we first notice that the Earth's tidal potential (Eq.~(\ref{potearth})) contains denominators depending on the radius $r_{\Earth}=(x_{\Earth}^2+y_{\Earth}^2+z_{\Earth}^2)^{1/2}$, and hence, depending on the trigonometric polynomials of Eqs.~(\ref{rE_new}). Through basic theory of the lunar orbit, one can check that the coefficients $C_{20}^{(x)}$, $C_{23}^{(y)}$ are of first order in the eccentricity $e_{\Moon}$ of the Moon's geocentric orbit, the coefficient $C_{23}^{(z)}$ is of first order in the inclination $i_{\Moon}$ of the Moon's geocentric orbit with respect to the ecliptic, while all remaining coefficients of the trigonometric polynomials (\ref{rE_new}) are of second order in the quantities $e_{\Moon},i_{\Moon}$. We then expand the potential term $\underline{V}_{\Earth}^{SSM}$ up to terms of second order in $e_{\Moon},i_{\Moon}$, arriving at
\begin{align*}
\underline{V}_{\Earth}^{SSM}(\ua,\ue,\ui,\ug,\uh,\varphi_1,\varphi_2,\varphi_3,\varphi_4)=
\left(\underline{V}_{\Earth}^{SSM, 0} + \underline{V}_{\Earth}^{SSM, 1}+\underline{V}_{\Earth}^{SSM, 2}\right)(\ua,\ue,\ui,\ug,\uh,\varphi_1,\varphi_2,\varphi_3,\varphi_4)+\Oscr(e_{\Moon}, i_{\Moon})^3\, .
\end{align*}
\noindent Re-arranging the terms $\underline{V}_{\Earth}^{SSM, 0}$, $\underline{V}_{\Earth}^{SSM, 1}$, $\underline{V}_{\Earth}^{SSM, 2}$ so to split their trigonometric part by the non-trigonometric one, we can write the potential as a truncated polynomial trigonometric expression of the form
\begin{equation}\label{potearthexp}
\underline{V}_{\Earth}^{SSM}(\ua,\ue,\ui,\ug,\uh,\varphi_1,\varphi_2,\varphi_3,\varphi_4)=
\underline{V}_{\Earth,0}^{SSM}(\ua,\ue,\ui)+
\underline{V}_{\Earth,1}^{SSM}(\ua,\ue,\ui,\ug,\uh,\varphi_1,\varphi_2,\varphi_3,\varphi_4)
\end{equation}
where
\begin{equation*}
\begin{aligned}
\underline{V}_{\Earth,1}^{SSM}(\ua,\ue,\ui,\ug,\uh,&\varphi_1,\varphi_2,\varphi_3,\varphi_4)=\\
&\sum_{\substack{n,m,l,s,k_1,k_2\\|n|+|m|+|l|\\+|s|+|k_1|+|k_2|\neq 0}}
c_{\Earth,n,m,l,s,k_1,k_2}(\ua,\ue,\ui)\cos(k_1g+k_2h+n\varphi_1+m\varphi_2+l\varphi_3+s\varphi_4)\\
+
&\sum_{\substack{n,m,l,s,k_1,k_2\\|n|+|m|+|l|\\+|s|+|k_1|+|k_2|\neq 0}}
d_{\Earth,n,m,l,s,k_1,k_2}(\ua,\ue,\ui)\sin(k_1g+k_2h+n\varphi_1+m\varphi_2+l\varphi_3+s\varphi_4)~~.
\end{aligned}
\end{equation*}
From closed form theory, we also have that 
\begin{equation}\label{potmoonexp}
\underline{V}_{\Moon}^{SSM}(\ua,\ue,\ui,\ug,\uh)=
\underline{V}_{\Moon,0}^{SSM}(\ua,\ue,\ui)+
\underline{V}_{\Moon,1}^{SSM}(\ua,\ue,\ui,\ug,\uh)
\end{equation}
where
\begin{equation*}
\begin{aligned}
\underline{V}_{\Moon,1}^{SSM}(\ua,\ue,\ui,\ug,\uh)&=\\
&\sum_{\substack{k_1,k_2\\|k_1|+|k_2|\neq 0}}
c_{\Moon,k_1,k_2}(\ua,\ue,\ui)\cos(k_1g+k_2h)\\
+
&\sum_{\substack{k_1,k_2\\|k_1|+|k_2|\neq 0}}
d_{\Moon,k_1,k_2}(\ua,\ue,\ui)\sin(k_1g+k_2h)~~.
\end{aligned}
\end{equation*}
Finally, the Hamiltonian with the potentials expanded as above can become formally autonomous by introducing dummy action variables $I_j$ conjugate to the angles $\varphi_j$, as well as the frequencies $\nu_1=\dot{\lambda}_{\Moon}$, $\nu_2=\dot{\varpi}_{\Moon}$, $\nu_3=\dot{h}_{\Moon}$, $\nu_4=\dot{\ell}_{\Sun}$. This leads to:
\begin{equation}\label{hamsm_simply}
\begin{aligned}
\Hscr_{sec}(\ua,\ue,\ui, I_1, I_2, I_3, I_4, \ug,\uh, \varphi_1, \varphi_2, \varphi_3, \varphi_4)
&=\Zscr_{sec}(\ua,\ue,\ui, I_1, I_2, I_3, I_4) \\
&+\Hscr_{sec,1}(\ua,\ue,\ui, I_1, I_2, I_3, I_4, \ug,\uh, \varphi_1, \varphi_2, \varphi_3, \varphi_4)
\end{aligned}
\end{equation}
where
\begin{equation}\label{nfsec}
\begin{aligned}
\Zscr_{sec}(\ua,\ue,\ui, I_1, I_2, I_3, I_4)
&=
-{\mathcal{G} M_{\Moon}\over 2\ua}-\nu_1 \uH
+ \nu_1 I_1+ \nu_2 I_2+ \nu_3 I_3+ \nu_4 I_4\\
&+\underline{V}_{\Moon,0}^{SSM}(\ua,\ue,\ui)
+\underline{V}_{\Earth,0}^{SSM}(\ua,\ue,\ui)
\end{aligned}
\end{equation}
while
\begin{equation}\label{hamsec1}
\begin{aligned}
\Hscr_{sec,1}(\ua,\ue,\ui,&\ug,\uh,\varphi_1,\varphi_2,\varphi_3,\varphi_4)
=
\underline{V}_{\Moon,1}^{SSM}(\ua,\ue,\ui,\ug,\uh)
+
\underline{V}_{\Earth,1}^{SSM}(\ua,\ue,\ui,\ug,\uh,\varphi_1,\varphi_2,\varphi_3,\varphi_4) \\
&=
\sum_{\substack{n,m,l,s,k_1,k_2\\|n|+|m|+|l|\\+|s|+|k_1|+|k_2|\neq 0}}
c_{n,m,l,s,k_1,k_2}(\ua,\ue,\ui)\cos(k_1g+k_2h+n\varphi_1+m\varphi_2+l\varphi_3+s\varphi_4)\\
&+
\sum_{\substack{n,m,l,s,k_1,k_2\\|n|+|m|+|l|\\+|s|+|k_1|+|k_2|\neq 0}}
d_{n,m,l,s,k_1,k_2}(\ua,\ue,\ui)\sin(k_1g+k_2h+n\varphi_1+m\varphi_2+l\varphi_3+s\varphi_4)~~.
\end{aligned}
\end{equation}

\section{Secular normal form and analytical solution}
\label{sec:analytical}

\subsection{Secular normal form}
\label{subsec:normal_form}
Consider the canonical transformation from the Delaunay variables $(\uL,\uG,\uH,\uell,\ug,\uh)$, associated with the mean elements, to new action-angle variables given by
\begin{equation}
\label{can_trasf}
\begin{aligned}
&\lambda=\lambda,&\,\, &\varphi_{\gamma}=\gamma-\varphi_1, &\,\, &\varphi_{\theta}=\theta-\varphi_1,&\,\, &\tilde{\varphi}_1 = \varphi_1,&\,\, &\varphi_2=\varphi_2,&\,\, &\varphi_3=\varphi_3,&\,\, &\varphi_4=\varphi_4,\\
&\Lambda=\Lambda, &\,\, & J_{\Gamma}=\Gamma, &\,\, &J_{\Theta}=\Theta, &\,\, &J_1=I_1 + \Gamma+\Theta, &\,\, &I_2=I_2, &\,\, &I_3=I_3, &\,\, &I_4=I_4\, ,
\end{aligned}
\end{equation}
where $(\Lambda,\Gamma,\Theta,\lambda,\gamma,\theta)$ are modified Delaunay variables defined as
\begin{equation} \label{delaunay_mod}
\Lambda = \uL, \quad \Gamma = \uL-\uG, \quad \Theta= \uG -\uH, \quad \lambda=\uell+\ug+\uh, \quad \gamma=-\ug-\uh, \quad \theta=-\uh~~.
\end{equation}
The transformation (\ref{can_trasf}) leads to the definition of two slow angles $(\varphi_{\gamma}, \varphi_{\theta})$ with secular frequencies depending linearly on the coefficients $C_{nm}$ and $S_{nm}$ of Eq.~(\ref{ssmset}) as well as on the coefficient $\cgrav M_{\Earth}/a_{\Moon}^3$ with $a_{\Moon}\simeq C_1^{(x)}=$~semi-major axis of the geocentric lunar orbit. Note that, due to the presence of the centrifugal term  $-\nu_1\uH\simeq-\omega_{\Moon,z}\uH$ in the Hamiltonian, the angles $(\gamma,\theta)$ are only semi-secular, since $\dot{\gamma}\simeq\omega_{\Moon,z}+\ldots$, $\dot{\theta}\simeq\omega_{\Moon,z}+\ldots$. In physical terms, the slow precession of the satellite's line of nodes in the Moon's equatorial plane appears in the PALRF frame as a precession nearly opposite to the Moon's (semi-slow) rotation.  

The actions $(\Lambda,J_\Gamma,J_\Theta)$ are still connected to the elements $(a,e,i)$ through the inverse of the functions $a(\Lambda)$, $e(\Lambda,\Gamma)$, $i(\Lambda,\Gamma,\Theta)$. The secular Hamiltonian takes now the form 
\begin{equation}\label{hammod}
\begin{aligned}
\Hscr_{sec}(\ua,\ue,\ui,J_1, I_2, I_3, I_4,\varphi_\gamma,\varphi_\theta,\tilde{\varphi}_1, \varphi_2, \varphi_3, \varphi_4)
&=\Zscr_{sec}(\ua,\ue,\ui,J_1, I_2, I_3, I_4) \\
&+\varepsilon 
\Hscr_{sec,1}(\ua,\ue,\ui,\varphi_\gamma,\varphi_\theta,\tilde{\varphi}_1,\varphi_2,\varphi_3,\varphi_4)
\end{aligned}
\end{equation}
where
\begin{equation}\label{nfsecmod}
\begin{aligned}
\Zscr_{sec}(\ua,\ue,\ui,J_1, I_2, I_3, I_4)
&=
-{\mathcal{G} M_{\Moon}\over 2\ua}-\nu_1 \Lambda+ \nu_1 J_1+ \nu_2 I_2+ \nu_3 I_3+ \nu_4 I_4\\
&+\underline{V}_{\Moon,0}^{SSM}(\ua,\ue,\ui)
+\underline{V}_{\Earth,0}^{SSM}(\ua,\ue,\ui)
\end{aligned}
\end{equation}
while
\begin{equation}\label{hamsec1mod}
\begin{aligned}
\Hscr_{sec,1}(\ua,\ue,\ui,&\varphi_\gamma,\varphi_\theta,\tilde{\varphi}_1, \varphi_2, \varphi_3, \varphi_4)
=
\underline{V}_{\Moon,1}^{SSM}(\ua,\ue,\ui,\varphi_\gamma,\varphi_\theta)
+
\underline{V}_{\Earth,1}^{SSM}(\ua,\ue,\ui,\varphi_\gamma,\varphi_\theta,\tilde{\varphi}_1,\varphi_2,\varphi_3,\varphi_4) \\
&=
\sum_{\substack{n',m,l,s,k_1',k_2'\\|n'|+|m|+|l|\\+|s|+|k_1'|+|k_2'|\neq 0}}
\tilde{c}_{n',m,l,s,k_1',k_2'}(\ua,\ue,\ui)\cos(k_1'\varphi_\gamma+k_2'\varphi_\theta+n'\tilde{\varphi}_1+m\varphi_2+l\varphi_3+s\varphi_4)\\
&+
\sum_{\substack{n',m,l,s,k_1',k_2'\\|n'|+|m|+|l|\\+|s|+|k_1'|+|k_2'|\neq 0}}
\tilde{d}_{n',m,l,s,k_1',k_2'}(\ua,\ue,\ui)\sin(k_1'\varphi_\gamma+k_2'\varphi_\theta+n'\tilde{\varphi}_1+m\varphi_2+l\varphi_3+s\varphi_4)
\end{aligned}
\end{equation}
with $\tilde{c}_{n',m,l,s,k_1',k_2'}=c_{n,m,l,s,k_1,k_2}$, $\tilde{d}_{n',m,l,s,k_1',k_2'}=d_{n,m,l,s,k_1,k_2}$ for the indices $k_1'=-k_1$, $k_2'=k_1-k_2$, $n'=n-k_2$. 

The `book-keeping' symbol $\varepsilon$ in Eq.~(\ref{hammod}) has numerical value $\varepsilon=1$ and is used to denote terms of `first order' in perturbation theory. We now seek to compute a \textit{secular normal form}, i.e., integrable approximation to the Hamiltonian (\ref{hammod}). To this end, consider a near-to-identity canonical transformation 
\begin{equation}\label{traproper}
\mathcal{X}_p:
\vet{x}\rightarrow\vet{x'}=\vet{x}+\mathcal{O}(\varepsilon)
\end{equation}
where
$$
\vet{x}=(\Lambda, J_\Gamma, J_\Theta, J_1, I_2, I_3,I_4,\varphi_\gamma,\varphi_\theta,\tilde{\varphi}_1,\varphi_2, \varphi_3, \varphi_4)
$$
and 
$$
\vet{x}' =(\Lambda', J_\Gamma',J_\Theta', J_1', I_2', I_3', I_4',\varphi_\gamma', \varphi_\theta',\tilde{\varphi}_1', \varphi_2', \varphi_3', \varphi_4')
$$ 
are respectively the `old' (called \textit{mean}) and the `new' (called \textit{proper}) variables, before and after the transformation. The transformation $\mathcal{X}_p$ is defined so that 
\begin{equation}\label{nfsecnew}
\Hscr_{sec}(\vet{x}')=\Zscr_{sec}(\vet{x}')+\mathcal{O}(\varepsilon^2)\, ,
\end{equation}
i.e., all the dependence on the canonical angles through the Hamiltonian terms $\Hscr_{sec,1}$ is eliminated when the Hamiltonian is expressed in the proper instead of the mean variables. By standard perturbation theory, the transformation $\mathcal{X}_p$ is computed through the Lie series formula:
\begin{equation}\label{liechip}
\mathcal{X}_p:~~~
\vet{x'}=\vet{x}-\Lscr_{\chi_p(\vet{x})}\vet{x}+\mathcal{O}(\varepsilon^2)
\end{equation}
where $\Lscr_{\chi_p}=\{\cdot,\chi_p\}$ is the Poisson bracket operator on functions depending on the variables $\vet{x}$, and $\chi_p$ is the Lie generating function. The latter satisfies the homological equation 
\begin{equation}
\label{homo}
L_{\chi_p} \Zscr_{sec} + \varepsilon \Hscr_{sec,1} = 0\, ,
\end{equation}
whose solution reads:
\begin{equation}\label{chip}
\begin{aligned}
\chi_p(\vet{x})&= \varepsilon\Big(
\sum_{\substack{n',m,l,s,k_1',k_2'\\
|n'|+|m|+|l|+|s|\\+|k_1'|+|k_2'|\neq 0} } 
-\frac{\tilde{d}_{n',m,l,s,k_1',k_2'}(\ua,\ue,\ui)}
{\mathscr{D}_{n',m,l,s,k_1',k_2'}} 
\cos(k_1'\varphi_\gamma+k_2'\varphi_\theta+n'\tilde{\varphi}_1+m\varphi_2+l\varphi_3+s\varphi_4) \\
&~+ 
\sum_{\substack{n',m,l,s,k_1',k_2'\\
|n'|+|m|+|l|+|s\\|+|k_1'|+|k_2'|\neq 0} } 
\frac{\tilde{c}_{n',m,l,s,k_1',k_2'}(\ua,\ue,\ui)}
{\mathscr{D}_{n',m,l,s,k_1',k_2'}} 
\sin(k_1'\varphi_\gamma+k_2'\varphi_\theta+n'\tilde{\varphi}_1+m\varphi_2+l\varphi_3+s\varphi_4)\Big)~~,
\end{aligned}
\end{equation}
with divisors
\begin{equation}\label{divisors}
\mathscr{D}_{n',m,l,s,k_1',k_2'}=
k_1'\nu_\gamma(\ua,\ue,\ui)+k_2'\nu_\theta(\ua,\ue,\ui)
+n'\nu_1+m\nu_2+l\nu_3+s\nu_4
\end{equation}
computed through the linear combinations of the frequencies $\nu_j$, $j=1,\ldots 4$ as well as of the secular frequencies
\begin{equation}\label{freq}
\nu_\gamma(\ua,\ue,\ui) = {\partial\Zscr_{sec}\over\partial J_\Gamma},~~
\nu_\theta(\ua,\ue,\ui) = {\partial\Zscr_{sec}\over\partial J_\Theta}~~.
\end{equation}

\subsection{Proper elements and analytical solution}
\label{subsec:solanal}
By the properties of Lie transformations, the transform inverse to Eq.~\eqref{liechip} is given by
\begin{equation}\label{liechipinv}
\mathcal{X}_p^{-1}:~~~
\vet{x}=\vet{x'}+\Lscr_{\chi_p(\vet{x'})}\vet{x'}+\mathcal{O}(\varepsilon^2)\, .
\end{equation}
From the definition of the Lie generating function $\chi_p$ as in Eq.~(\ref{chip}) we then have the following properties:

i) The angles $(\tilde{\varphi}_1,\varphi_2,\varphi_3,\varphi_4)$ do not transform, i.e., the identity $(\tilde{\varphi}_1',\varphi_2',\varphi_3',\varphi_4')=(\tilde{\varphi}_1,\varphi_2,\varphi_3,\varphi_4)$ holds. This follows from the functional independence of $\chi_p$ on the dummy actions, and it is a consequence of the fact that these angles express the explicit dependence of the equations of motion on time. 

ii) The action $\Lambda$ does not transform neither ($\Lambda'=\Lambda$), since the Hamiltonian and generating function have no dependence on the angle $\lambda$ (mean longitude) conjugate to $\Lambda$. Physically, the mean semi-major axis $\ua$ is already a preserved quantity in the secular Hamiltonian before the normalization.   

iii) The dummy actions $(J_1,I_2,I_3,I_4)$ do not appear in the term $\Lscr_{\chi_p(\vet{x})}\vet{x}$ of the r.h.s. of Eq.~(\ref{liechip}), or the term $\Lscr_{\chi_p(\vet{x'})}\vet{x'}$ of the r.h.s. of Eq.~(\ref{liechipinv}). This implies that the forward and backward transformations of the action variables $(\Lambda,J_\Gamma=\Gamma,J_\Theta=\Theta)$ involve only the same set of three orbital action variables, while it involves the entire set of the canonical angles $(\varphi_\gamma,\varphi_\theta,\tilde{\varphi}_1,\varphi_2,\varphi_3,\varphi_4)$, or, equivalently, $(g,h,\varphi_1,\varphi_2,\varphi_3,\varphi_4)$.

iv) Ignoring the $\mathcal{O}(\varepsilon^2)$ terms in Eq.~(\ref{nfsecnew}), the secular Hamiltonian expressed in the new variables has no dependence on any of the canonical angles. Hence, the quantities $(\Lambda',J_\Gamma'=\Gamma',J_\Theta'=\Theta')$ are integrals of motion. The constant quantities $(a_p,e_p,i_p)$ with 
\begin{equation}\label{properel}
a_p = {\Lambda'^2\over\cgrav M_{\Moon}}=\ua~,~~~
e_p = \left(1-(1-\Gamma'/\Lambda')^2\right)^{1/2}~,~~~
\sin(i_p/2) = \left({\Theta'\over2(\Lambda'-\Gamma')}\right)^{1/2}
\end{equation}
are the proper semi-major axis, eccentricity and inclination of an orbit. 

To compute an analytical solution to the secular equations of motion for any given initial condition $(\ua(t_0),\ue(t_0),\ui(t_0),\uell(t_0),\ug(t_0),\uh(t_0))$, we first use the set of equations (\ref{can_trasf}), (\ref{liechip}), (\ref{properel}) to compute the initial conditions in the new variables $(a'(t_0)=a_p,e'(t_0)=e_p,i'(t_0)=i_p,\lambda'(t_0),\varphi_\gamma'(t_0),\varphi_\theta'(t_0))$ at the initial time $t_0$. We then obtain the analytical solution in all proper elements:
\begin{equation}\label{propersol}
\begin{aligned}
a'(t)&=a_p\,, \\
e'(t)&=e_p\,, \\
i'(t)&=i_p\,, \\
\lambda'(t)&=\lambda'(t_0)+\nu_\lambda(a_p,e_p,i_p) (t-t_0)\,, \\
\varphi_\gamma'(t)&=\varphi_\gamma'(t_0)+\nu_\gamma(a_p,e_p,i_p) (t-t_0)\,, \\
\varphi_\theta'(t)&=\varphi_\theta'(t_0)+\nu_\theta(a_p,e_p,i_p) (t-t_0)\, ,
\end{aligned}
\end{equation}
where
\begin{equation}\label{nulambda}
\nu_{\lambda}(a_p,e_p,i_p)=\left({\partial\Zscr_{sec}\over\partial\Lambda}\right)_{\ua=a_p,\ue=e_p,\ui=i_p}
\end{equation}
and the frequencies $\nu_\gamma(a_p,e_p,i_p)$, $\nu_\theta(a_p,e_p,i_p)$ are computed as in Eq.~(\ref{freq}), with the substitution $(\ua=a_p,\ue=e_p,\ui=i_p)$. The dummy actions $(J_1',I_2',I_3',I_4')$ are also integrals of motion, and, by property (iii) above, they can be assigned constant-in-time arbitrary values (e.g. all equal to zero). Together with Eqs.~\eqref{propersol},~\eqref{phases} and property (i) above, this specifies the evolution in time of the complete vector of proper variables $\vet{x}'(t)$. Using then the inverse transformation (\ref{liechipinv}) we obtain the evolution in time of the vector $\vet{x}(t)$, hence of the mean elements $(\ua(t),\ue(t),\ui(t),\uell(t),\ug(t),\uh(t))$. 

In practice, to avoid apparent singularities in the transformations~\eqref{liechip} or~\eqref{liechipinv}, we work with an alternative set of variables equivalent to equinoctial elements, and use the Poisson algebra structure reported in~\cite{eftetal2023} to compute all Poisson brackets in closed form with computer-algebraic functions acting directly on the elements instead of the canonical action-angle variables of Eq.~(\ref{can_trasf}). The complete procedure to obtain an analytic propagator, starting from the initial conditions of an orbit in osculating elements, is described in the next section. 

\section{Structure of the analytical propagator. Secular resonances 
and precision tests}
\label{sec:propagator}

\subsection{Computer-algebraic implementation}
\label{subsec:symbolic}
The complete analytic propagator combines the canonical transformations $\mathcal{X}$ (Eq.~(\ref{tramean})) and $\mathcal{X}_p$ (Eq.~\eqref{liechip}) to transform any initial datum $\zeta(t_0)$, given in Cartesian coordinates or osculating Keplerian elements at the initial time $t_0$, to the corresponding proper elements $(a_p,e_p,i_p)$ (constant), as well as $(\lambda'(t_0),\varphi_\gamma'(t_0),\varphi_\theta'(t_0))$, or, equivalently $(\ell'(t_0),g'(t_0),h'(t_0))$, with $\ell'=\lambda'+\varphi_\gamma'+\tilde{\varphi}_1$, $g'=\varphi_\theta'-\varphi_\gamma'$, $h'=-\varphi_\theta'-\tilde{\varphi}_1$. Then, the equations (\ref{propersol}), combined with (\ref{phases}), are used to express the analytical solution in time for all the proper variables. Finally, the inverse transformation $\mathcal{X}_p^{-1}$ (Eq.~\eqref{liechipinv}) is first used to compute the analytical solution in the mean elements $(\ua,\ue,\ui,\uell,\ug,\uh)$, and the transformation $\mathcal{X}^{-1}$ is then used to compute the analytical solution in the osculating  elements $(a,e,i,\ell,g,h)$. The process requires no numerical propagation of any variable at any step of the algorithm, hence, in the corresponding routines one can connect immediately the osculating elements of an orbit at two different times $(t_0,t)$
$$
(a(t_0),e(t_0),i(t_0),\ell(t_0),g(t_0),h(t_0))
\rightarrow
(a(t),e(t),i(t),\ell(t),g(t),h(t))
$$
with $t$ in the past or future of $t_0$, without any numerical propagation between the times $[t_0,t]$. This is an essential advantage over numerical propagation of the trajectories, since it practically reduces to zero the computational cost involved in the numerical integration in the interval $[t_0,t]$. 

Note also that both transformations $\mathcal{X}$, $\mathcal{X}_p$ are subject to the well known apparent singularities associated with orbits crossing the Moon's equator ($i=0$), or becoming instantaneously circular ($e=0$), since the angles $g$ and $h$ are not possible to define in these cases. However, all apparent singularities are removed when the transformations are expressed in the set of non-singular variables
\begin{equation}
\begin{aligned}
& h_{eq} = e\cos(\varpi) = e \cos(g+h), &\qquad & k_{eq} = e\sin(\varpi) = e \sin(g+h),\\
& q_{eq}= \sin (i/2) \cos(h), &\qquad & p_{eq}= \sin (i/2) \sin(h).
\end{aligned}
\end{equation}
The set $(\lambda, k_{eq}, p_{eq}, a, h_{eq}, q_{eq})$ are called \textit{modified equinoctial variables}, and they can be distinguished to osculating, mean, or proper, depending on $(\ell, g, h, a, e, i)$, $(\uell, \ug, \uh, \ua, \ue, \ui)$, or 
$(\ell', g', h', a'=a_p, e'=e_p, i'=i_p)$ respectively. In particular, the transformation $\mathcal{X}_p$ of Eq.~(\ref{liechip}) for the passage from mean to proper modified equinoctial elements takes the form:
\begin{align*}
h'_{eq}&= \left( e\cos(g+h)-\poisson{e\cos(g+h)}{\chi_p}\right)_{a=\ua,e=\ue,i=\ui,\ell=\uell,g=\ug,h=\uh}, \\
k'_{eq}&= \left( e\sin(g+h)-\poisson{e\sin(g+h)}{\chi_p}\right)_{a=\ua,e=\ue,i=\ui,\ell=\uell,g=\ug,h=\uh}, \\
q'_{eq}&= \left( \sin(i/2)\cos(h)-\poisson{\sin(i/2)\cos(h)}{\chi_p}\right)_{a=\ua,e=\ue,i=\ui,\ell=\uell,g=\ug,h=\uh}, \\
p'_{eq}&= \left( \sin(i/2)\sin(h)-\poisson{\sin(i/2)\sin(h)}{\chi_p}\right)_{a=\ua,e=\ue,i=\ui,\ell=\uell,g=\ug,h=\uh},
\end{align*}
where all Poisson brackets $\{\cdot,\cdot\}$ are computed through a computer-algebraic manipulator using the same definitions and simplification rules for the Poisson algebra as in the semi-analytical propagator SELENA (\cite{eftetal2023}). The inverse transformation $\mathcal{X}_p^{-1}$ is computed in the same way, leading to
\begin{align*}
\underline{h}_{eq}&= 
\left( e\cos(g+h)+\poisson{e\cos(g+h)}{\chi_p}\right)_{a=a',e=e',i=i',\ell=\ell',g=g',h=h'}, \\
\underline{k}_{eq}&= 
\left( e\sin(g+h)+\poisson{e\sin(g+h)}{\chi_p}\right)_{a=a',e=e',i=i',\ell=\ell',g=g',h=h'}, \\
\underline{q}_{eq}&= 
\left( \sin(i/2)\cos(h)+\poisson{\sin(i/2)\cos(h)}{\chi_p}\right)_{a=a',e=e',i=i',\ell=\ell',g=g',h=h'}, \\
\underline{p}_{eq}&= 
\left( \sin(i/2)\sin(h)+\poisson{\sin(i/2)\sin(h)}{\chi_p}\right)_{a=a',e=e',i=i',\ell=\ell',g=g',h=h'}~~,
\end{align*}
while for the transformation $\mathcal{X}$ from osculating to mean equinoctial elements and its inverse we reproduce exactly the formulas reported in~\cite{eftetal2023}. 

\subsection{Accuracy tests of the solution}
\begin{figure}[!h]
\centering
\includegraphics[scale=0.5]{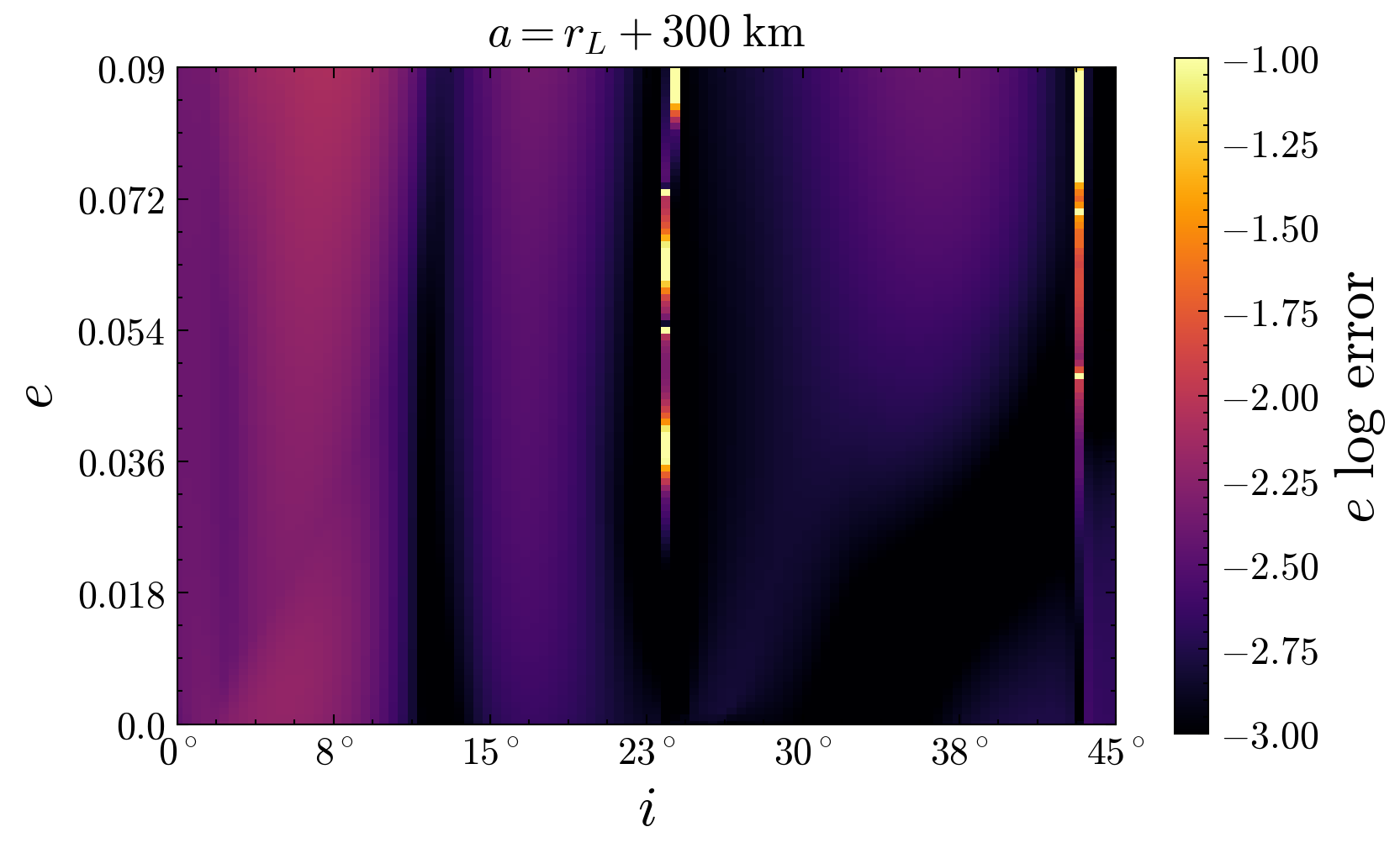}
\includegraphics[scale=0.5]{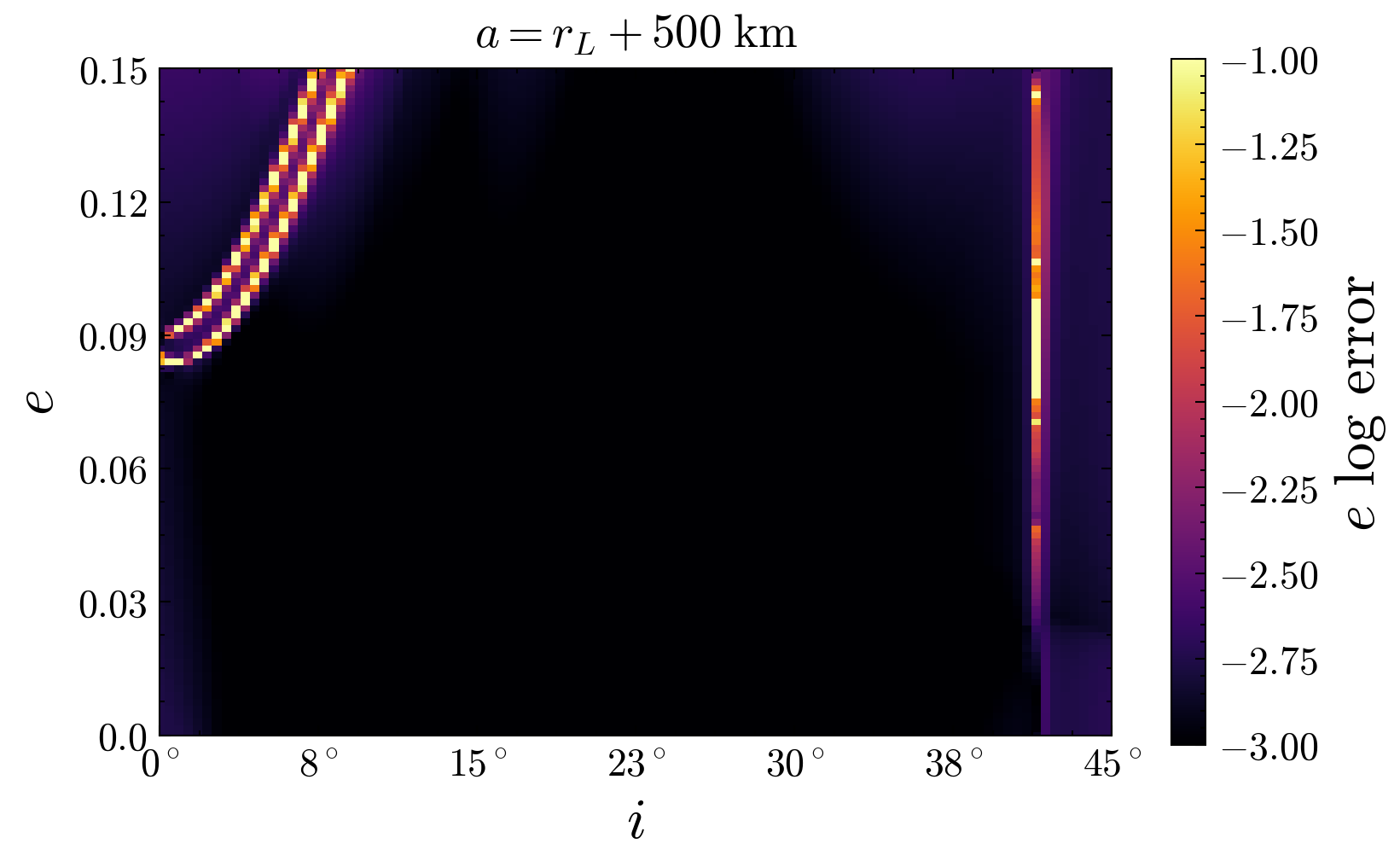}
\includegraphics[scale=0.5]{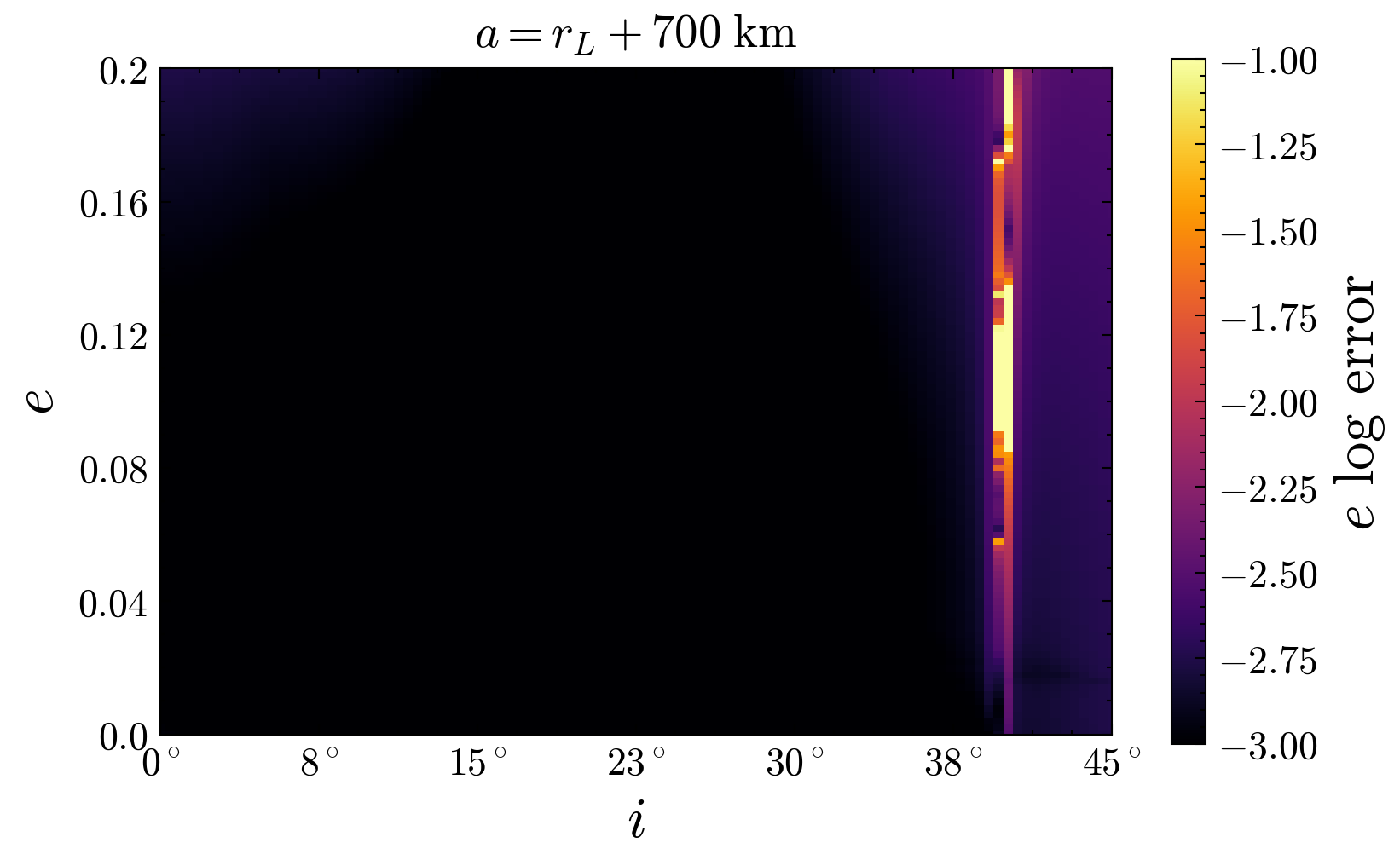}
\includegraphics[scale=0.5]{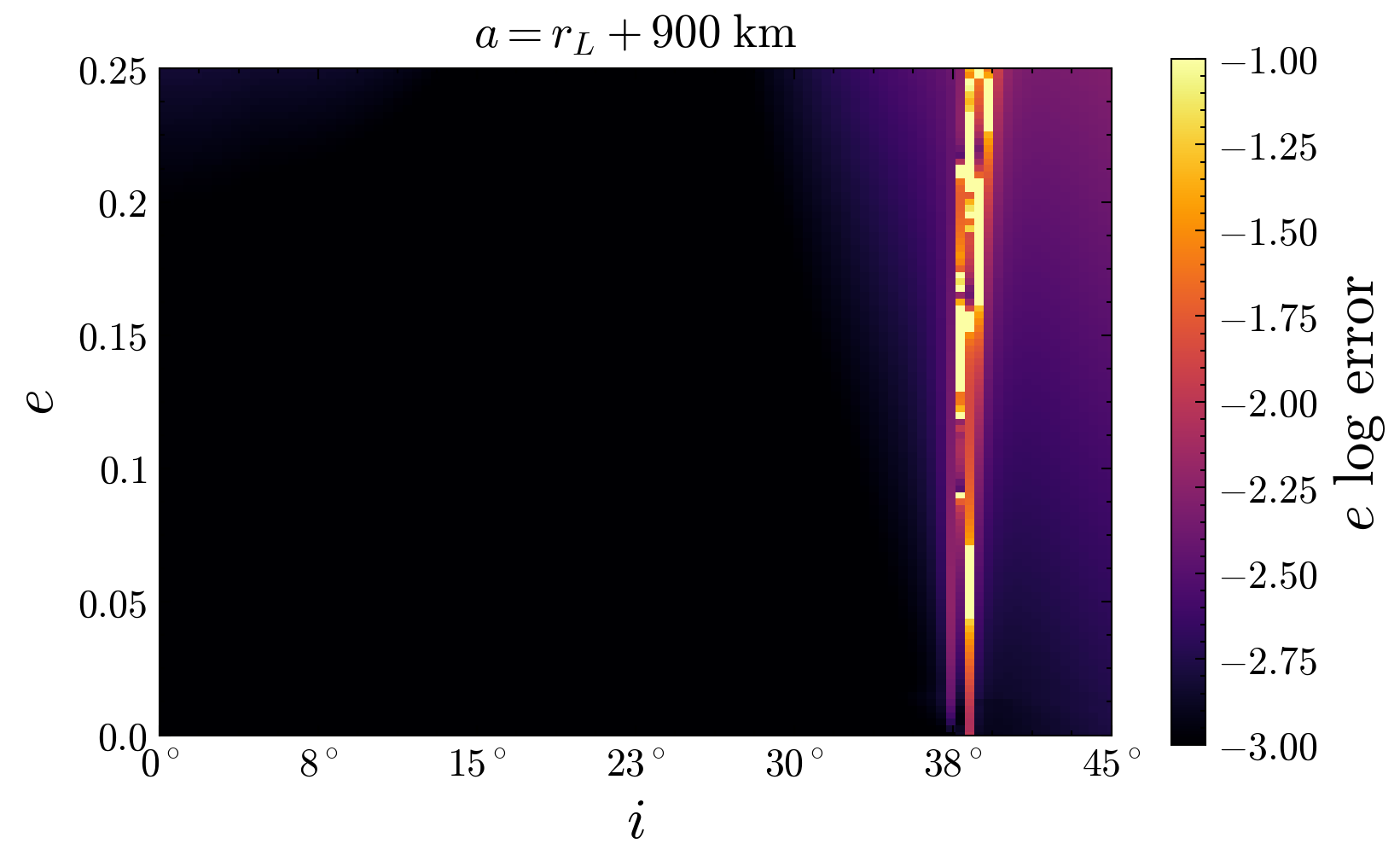}
\includegraphics[scale=0.5]{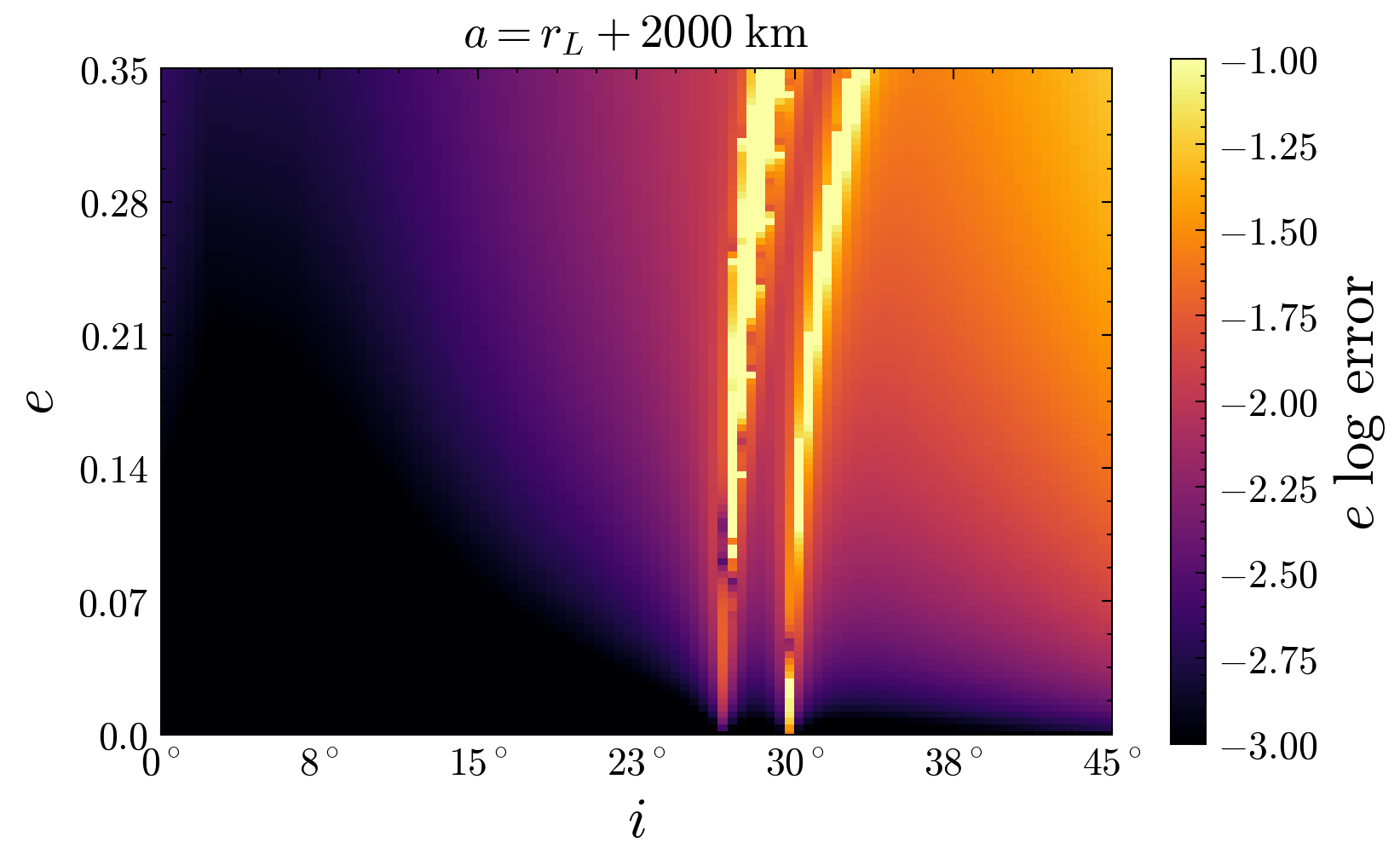}
\includegraphics[scale=0.5]{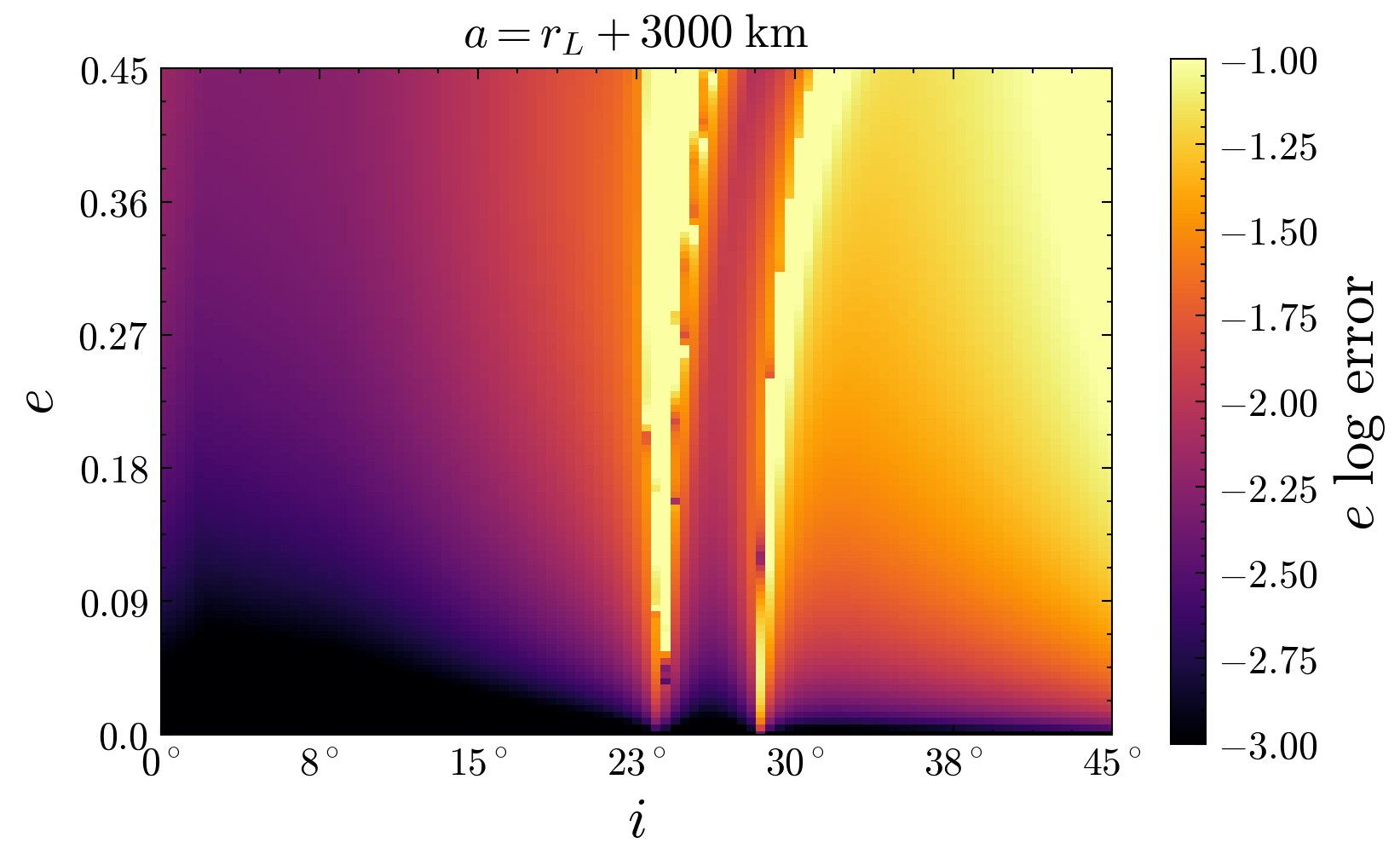}
\caption{
\small Maps of the maximum value of $\log_{10}(|\Delta\ue(t)|)$ in the timespan $0\leq t\leq 178$~days, where $\Delta\ue(t)=\ue_a(t)-\ue_n(t)$, with $\ue_{a}(t)=$~analytically propagated, and $\ue_{n}(t)=$~numerically propagated (with the Hamiltonian $\Hscr_{sec}$, Eq.~\eqref{hamsm_simply}) mean eccentricity of an orbit. The value of $\log_{10}(|\Delta\ue(t)|)$ is given in color scale as indicated in each figure. The map is computed over a $100\times 100$ grid of initial conditions in $\ue(0),\ui(0)$ within the limits indicated in each panel, with the remaining mean elements being $\uell(0)=0$, $\ug(0)=-0.4$ rad, $\uh(0)= 0.7$ rad, and $\ua(0)=a_p=const=R_{\Moon}+\delta a$, where $\delta a = (300,500,700,900,2000,3000)~$km. Orbits with error smaller than $10^{-3}$ are classified as of error $10^{-3}$, while those with error larger than $10^{-1}$ are classified as of error $10^{-1}$. 
}
\label{fig:errormapsecc}
\end{figure}

\begin{figure}[!h]
\centering
\includegraphics[scale=0.5]{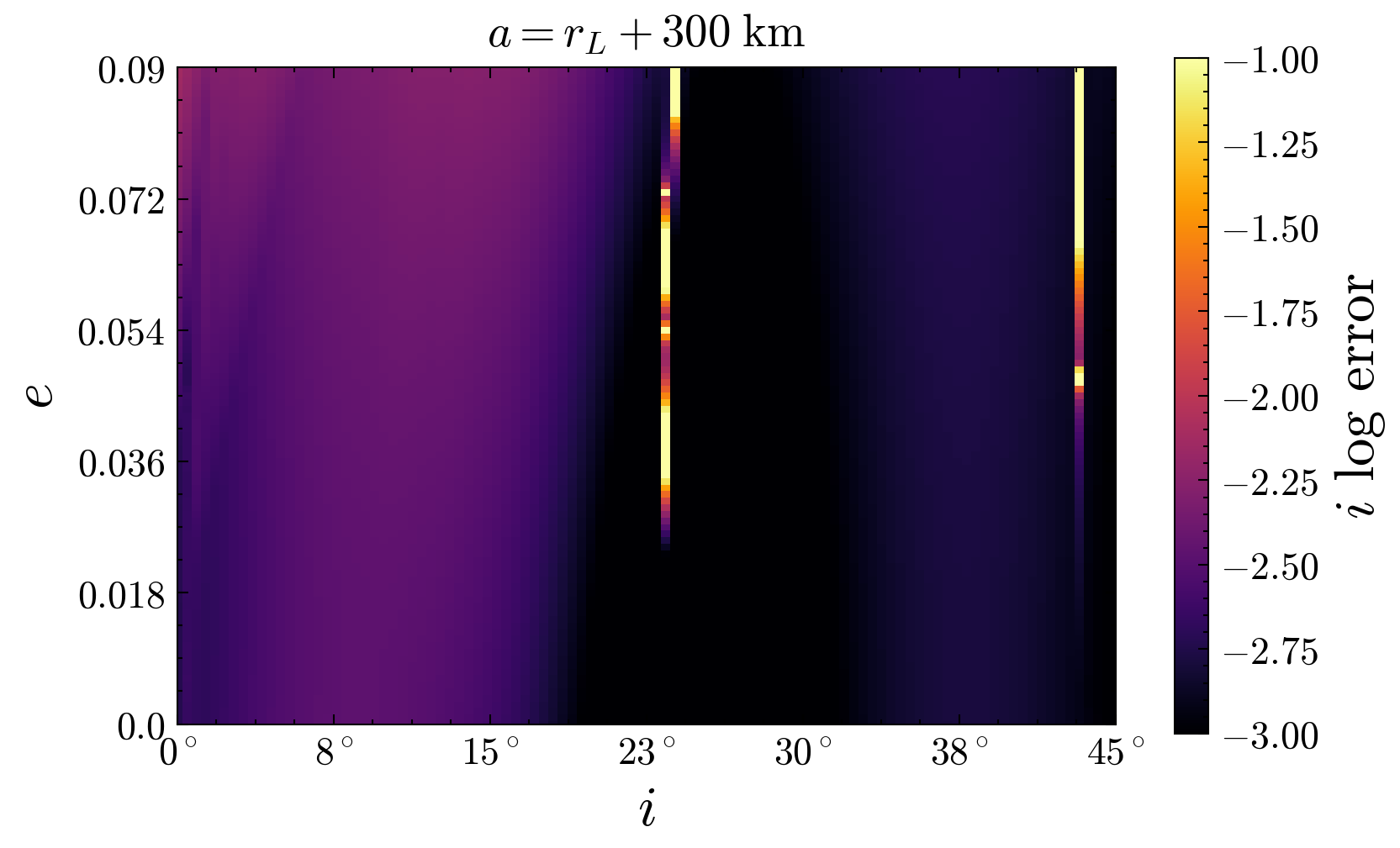}
\includegraphics[scale=0.5]{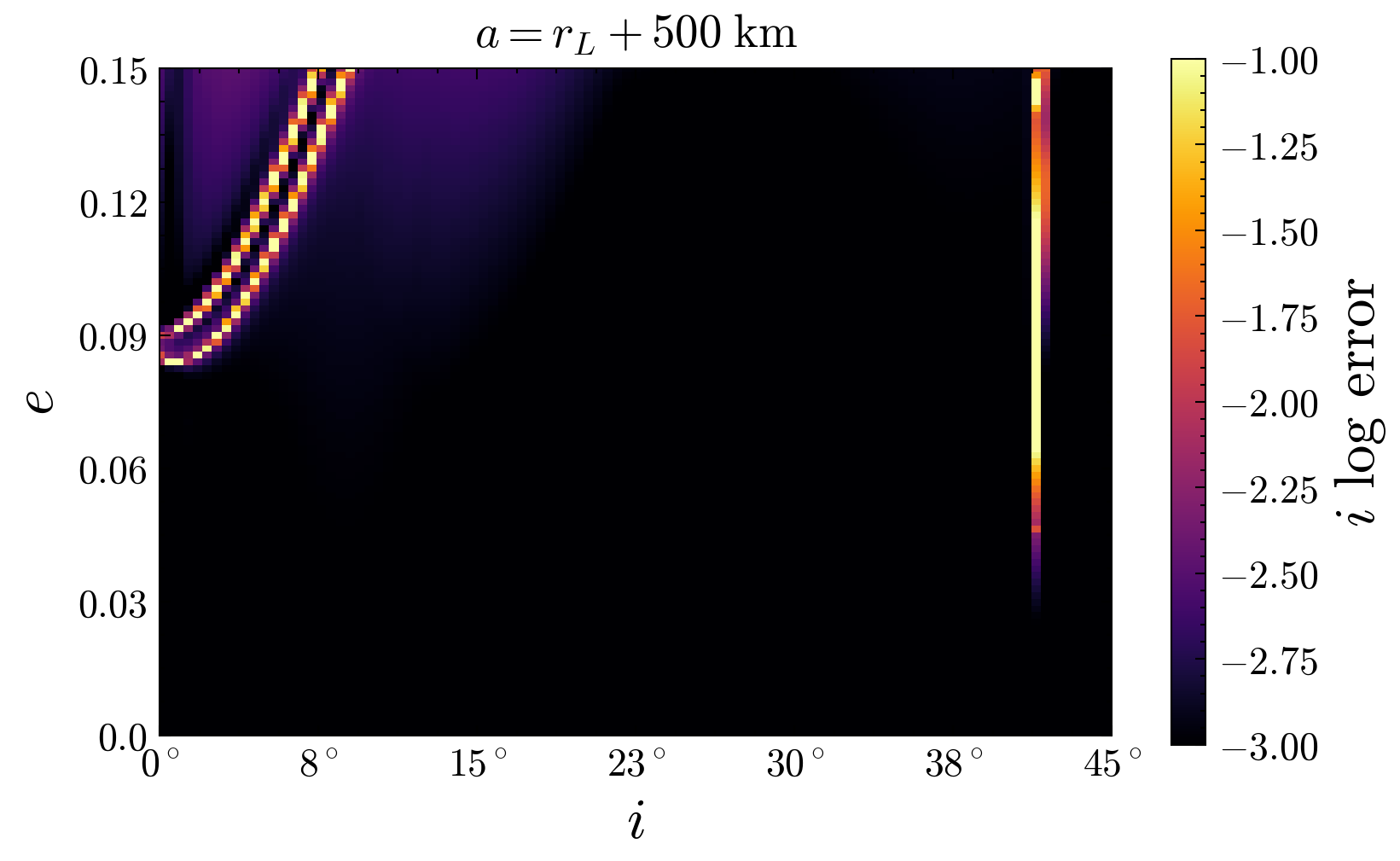}
\includegraphics[scale=0.5]{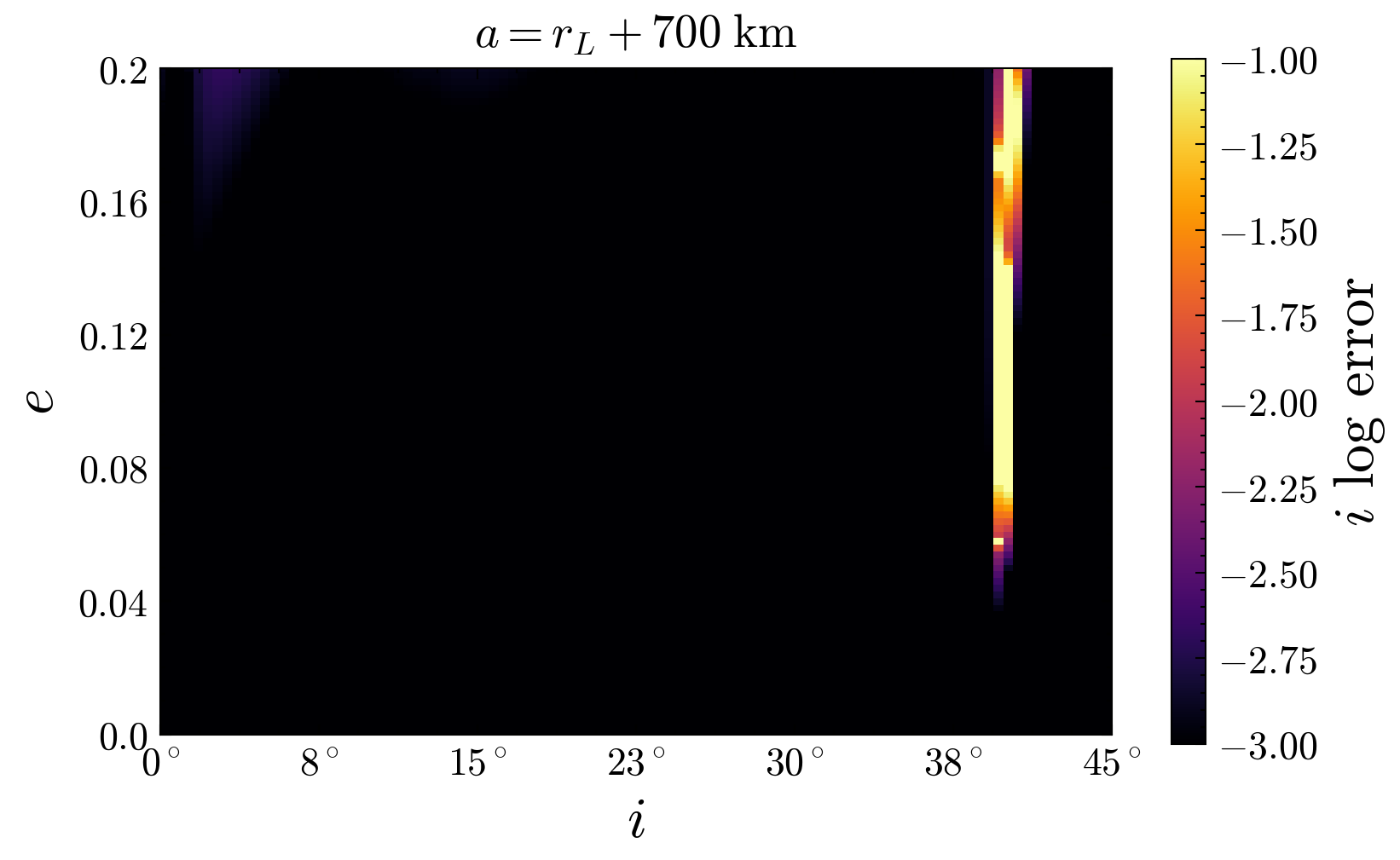}
\includegraphics[scale=0.5]{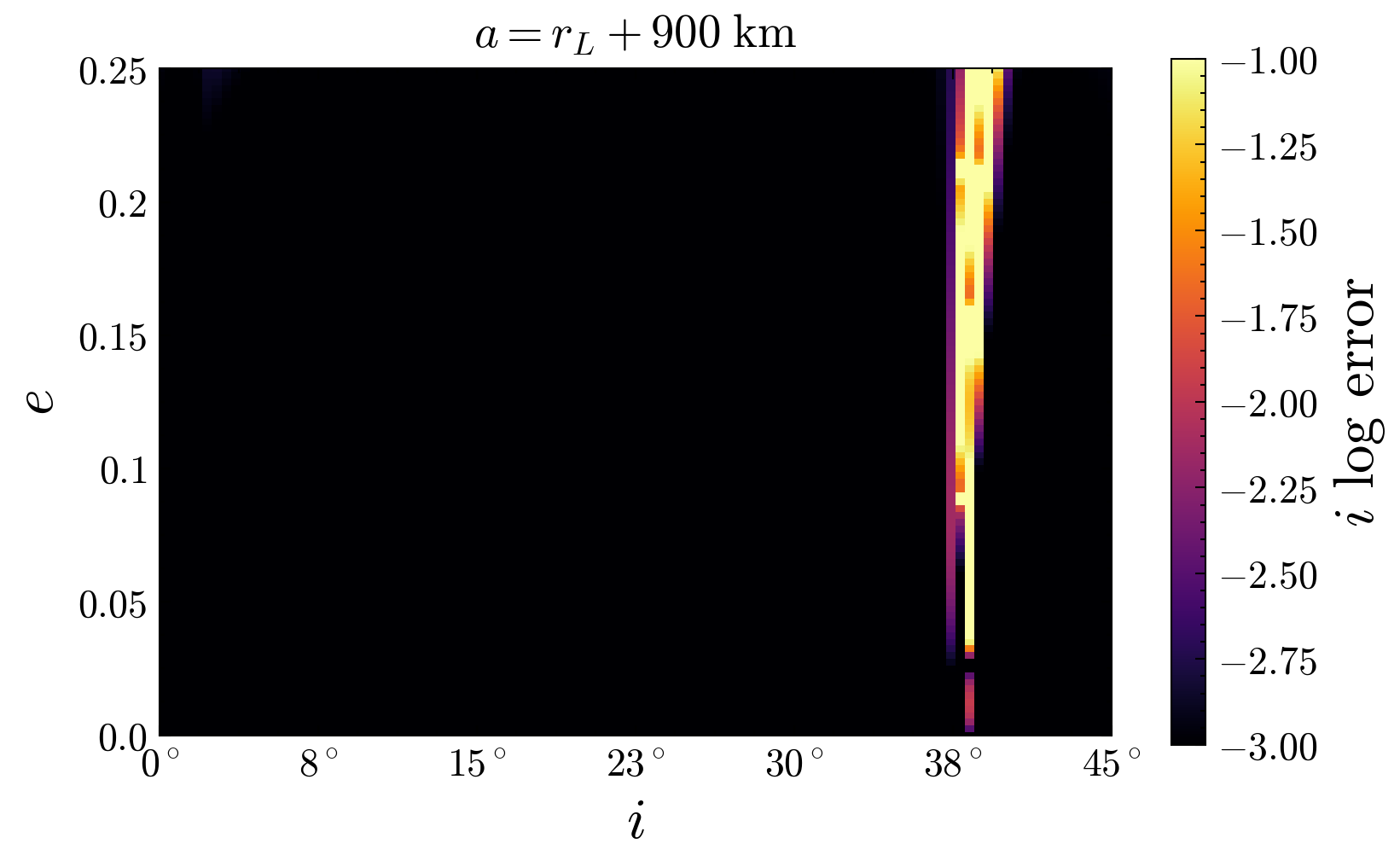}
\includegraphics[scale=0.5]{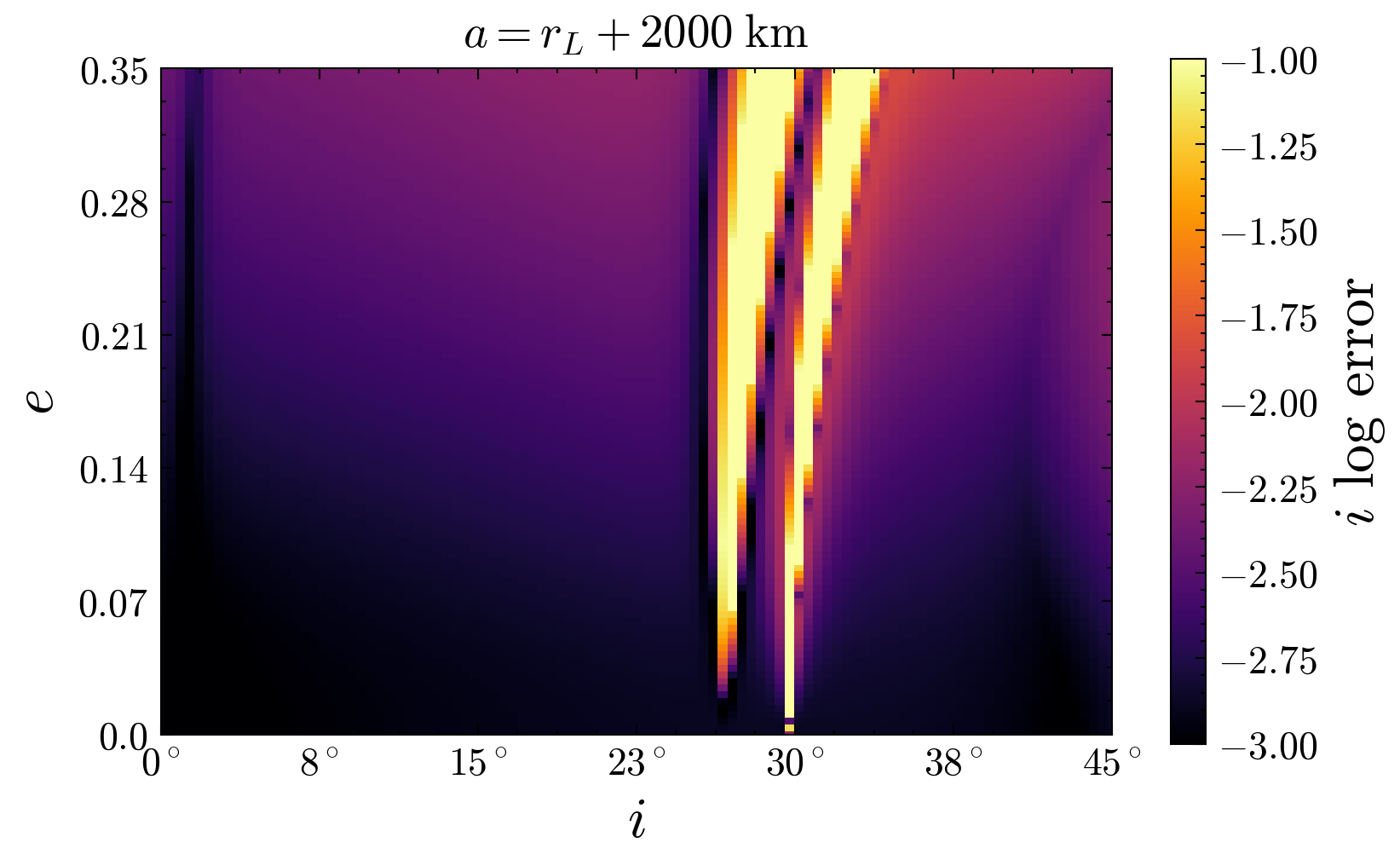}
\includegraphics[scale=0.5]{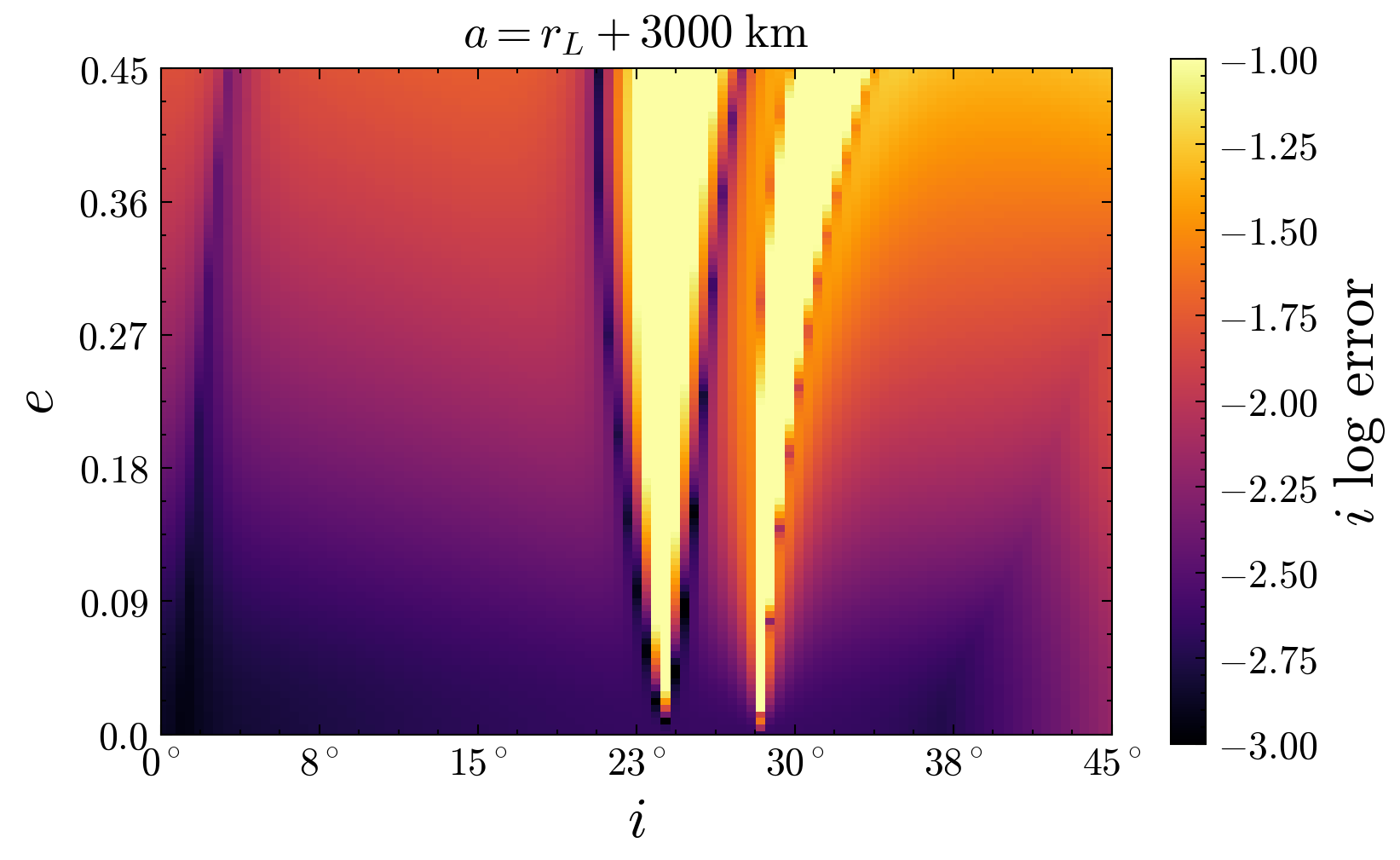}
\caption{
\small Same as in Fig.~\ref{fig:errormapsecc}, but for the maximum error $\Delta\ui(t)=\ui_a(t)-\ui_n(t)$ in the analytical versus numerical propagation of the mean inclination of the trajectories. The error is given in radians, hence the allowance to use a logarithmic color scale.  
}
\label{fig:errormapsinc}
\end{figure}

Errors in the analytical propagation of the trajectories, compared with numerical propagation in a fully Cartesian model, come from three sources, of which the two first were discussed in \cite{eftetal2023} and~\cite{legeft2024}, while the third is a subject of focus in the present section. These sources are: \\
\\
\noindent
(i) The accuracy of the transformation $\mathcal{X}$ (Eq.~\eqref{tramean}) which connects osculating with mean elements along a satellite trajectory. This is used already in the semi-analytical theory to connect numerically evolved (by the secular Hamiltonian) mean elements, with their (analytically mapped through $\mathcal{X}^{-1}$) corresponding osculating elements. The error of the `semi-analytical' theory is discussed extensively in section 5 of \cite{eftetal2023}. The error is computed with respect to a fully Cartesian numerical propagation of trajectories in the complete $10\times 10$ lunar gravity model including also the Earth $P_2,P_3,P_2^2,P_4$ terms, and the Sun $P_2$ and SRP terms, called the `SELENA model'. It is found that for all trajectories the error grows linearly with time, mostly as a result of a $\mathcal{O}(10^{-6})$ error in the analytical evaluation of the frequency $\dot{\ell}$. The latter leads to a cumulative error of the order $\sim 1$--$10$ degrees in the value of the fast angle $\ell$ for most satellite trajectories after one year. Translated to Cartesian coordinates, this gives a cumulative error of $~10$--$100$~km in the position of a satellite as predicted semi-analytically after one year. \\
\noindent
(ii) The error associated with truncating the SELENA model to the `Simplified for Secular Dynamics Model' (SSM) introduced in Section 3 of \cite{legeft2024} and in subsection \ref{subsec:potential} above. From figure 7 of \cite{eftetal2023}, as well as Figures 1, 7 and 8 of \cite{legeft2024} and accompanying discussion in the text, we have that the simplification in the SSM model with respect to the complete SELENA model yields an error whose estimate for $~1$~yr propagation is at the level of $10^{-4}$--$10^{-3}$ for most trajectories. Further evidence on this error is given below. \\
\noindent
(iii) The accuracy of the transformation $\mathcal{X}_p$ (Eq.~\eqref{liechip}), its inverse $\mathcal{X}_p^{-1}$ (which connect mean with proper elements along a trajectory) and of the analytical solution (\ref{propersol}) introduced in the present paper. \\
\\
We now focus on estimating the error (iii) and understanding its origin. Figures \ref{fig:errormapsecc} and \ref{fig:errormapsinc} provide the main information. Figure \ref{fig:errormapsecc} stems from a comparison of the evolution in time of the mean eccentricity $\ue(t)$ of a trajectory as propagated analytically (namely $\ue_a(t)$), using the formulas of our propagator, and numerically ($\ue_n(t))$, using Hamilton's equations
\begin{equation}\label{eqmoequi}
\begin{aligned}
\dot{\underline{h}_{eq}}&=\{\ue\cos(\ug+\uh),\Hscr_{sec}\} \\
\dot{\underline{k}_{eq}}&=\{\ue\sin(\ug+\uh),\Hscr_{sec}\} \\
\dot{\underline{q}_{eq}}&=\{\sin(\ui/2)\cos(\uh),\Hscr_{sec}\} \\
\dot{\underline{p}_{eq}}&=\{\sin(\ui/2)\sin(\uh),\Hscr_{sec}\}
\end{aligned}
\end{equation}
where $\Hscr_{sec}$ is the secular Hamiltonian model (\ref{hamsm_simply}) before any normalization. The Poisson brackets in Eq.~\eqref{eqmoequi} are computed with the same computer-algebraic representation as for the canonical transformations discussed in subsection~\ref{subsec:symbolic}.

Figure \ref{fig:errormapsecc} shows in color map the maximum value of the quantity $\log_{10}(|\Delta\ue(t)|)$ in the timespan $\mathcal{T}_{HY}:~0\leq t\leq 178$~days~$=$~half-year, where the difference between analytical and numerical is set equal to $\Delta\ue(t)=\ue_a(t)-\ue_n(t)$. Since the eccentricity represents the percentual difference between the distance of the apsides and the value of the semi-major axis of a trajectory, the use of logarithm directly represents percentual errors in the trajectory in the usual orbital configuration space. The quantity $\max_{\mathcal{T}_{HY}}\log_{10}(|\Delta\ue(t)|)$ is computed for all trajectories with fixed initial values of the mean angles $\uell(0)=0$, $\ug(0)=-0.4$ rad, $\uh(0)=0.7$ rad, and (constant) mean semi-major axis $\ua=R_{\Moon}+\delta a$, where $\delta a$ (the `altitude') is altered in the eight panels of Fig.~\ref{fig:errormapsecc} to the values $\delta a[$km$]=300$, $500$, $700$, $900$, $2000$, $3000$. These values of the semi-major axis are chosen so as to obtain figures which can be compared and complement the information given in Figure 5 of \cite{legeft2024}. For each choice of value of the semi-major axis, we compute the error in a $100\times 100$ grid of initial conditions for $(\ue,\ui)$ with $0\leq\ue(0)\leq 0.7\delta a/\ua$, i.e., about $70\%$ of the value of the maximum eccentricity for which the orbit's initial pericenter is above the Moon's surface, and $0^\circ\leq\ui(0)\leq 45^\circ$. The choice of these limits is motivated by visual inspection of Figure 5 of \cite{legeft2024}, which shows that the above limits essentially contain the entire domain of trajectories whose initial conditions $(\ua,\ue(0),\ui(0))$ do not lead to collision with the Moon's surface. Also, Figure \ref{fig:errormapsinc} shows in color map the maximum difference in mean inclination $\log_{10}(|\Delta\ui(t)|)$ of the analytical from the numerical solution for the same trajectories. 

\begin{figure}
\centering
\includegraphics[scale=0.55]{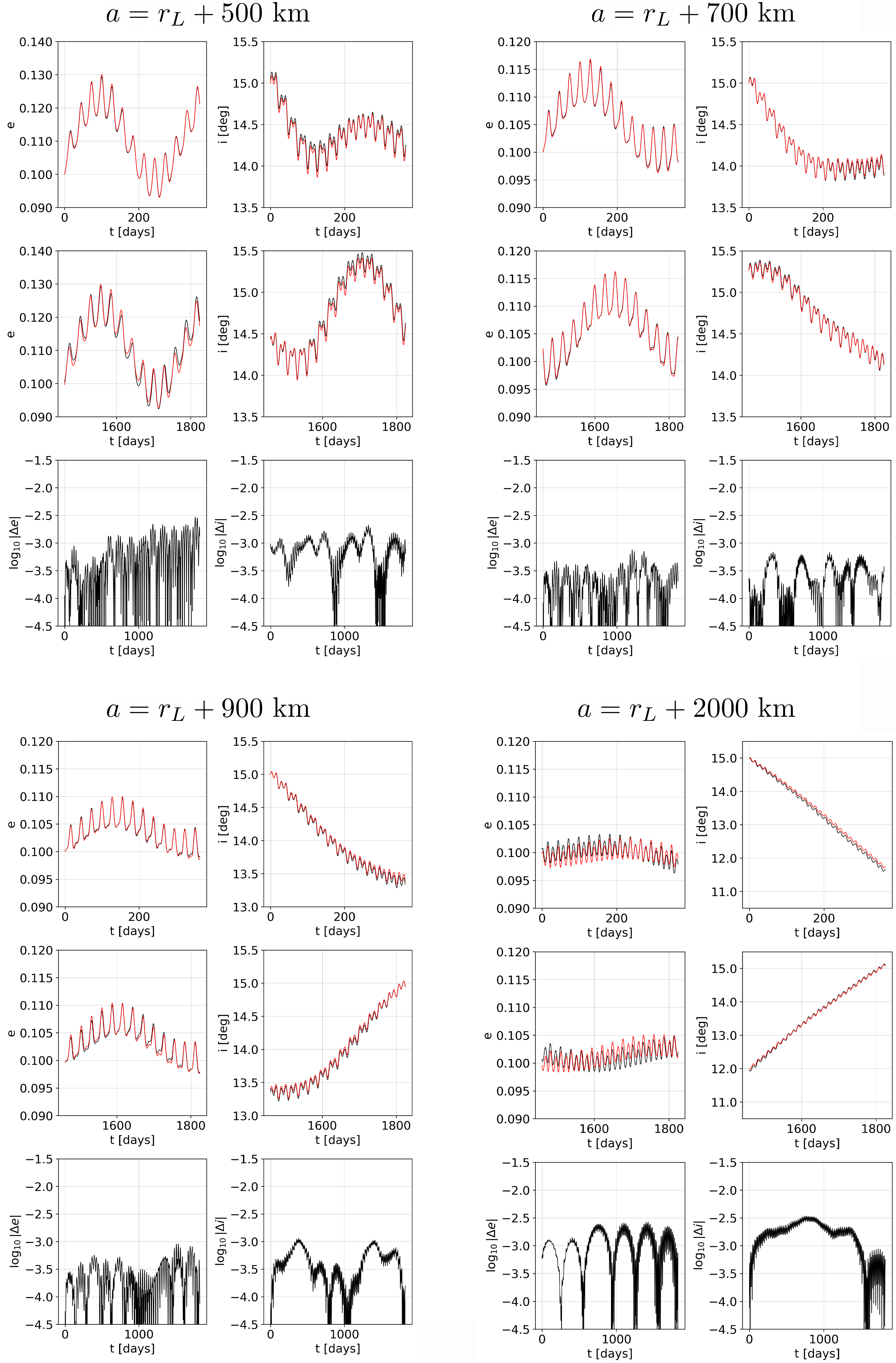}
\caption{\small   
Analytical (black line) versus numerical (red line) propagation of the mean elements $(\ue(t),\ui(t))$ for four orbits with initial conditions $\uell(0)=0$, $\ug(0)=-0.4$~rad, $\uh(0)=0.7$~rad, $\ue(0)=0.1$, $\ui(0)=15^\circ$ and (constant) semi-major axis corresponding to the group of panels $\ua=R_{\Moon}+500$~km (top left), $\ua=R_{\Moon}+700$~km (top right), $\ua=R_{\Moon}+900$~km (bottom left), $\ua=R_{\Moon}+2000$~km (bottom right). In each group of panels, the first row shows the evolution of the corresponding elements within the first year of propagation, the second row within the fifth year of propagation, and the third row shows the corresponding errors of the comparison of the analytical and numerical solution in $\log_{10}$ scale. 
}
\label{fig:orbitsgood}
\end{figure}
\begin{figure}[!h]
\centering
\includegraphics[scale=0.55]{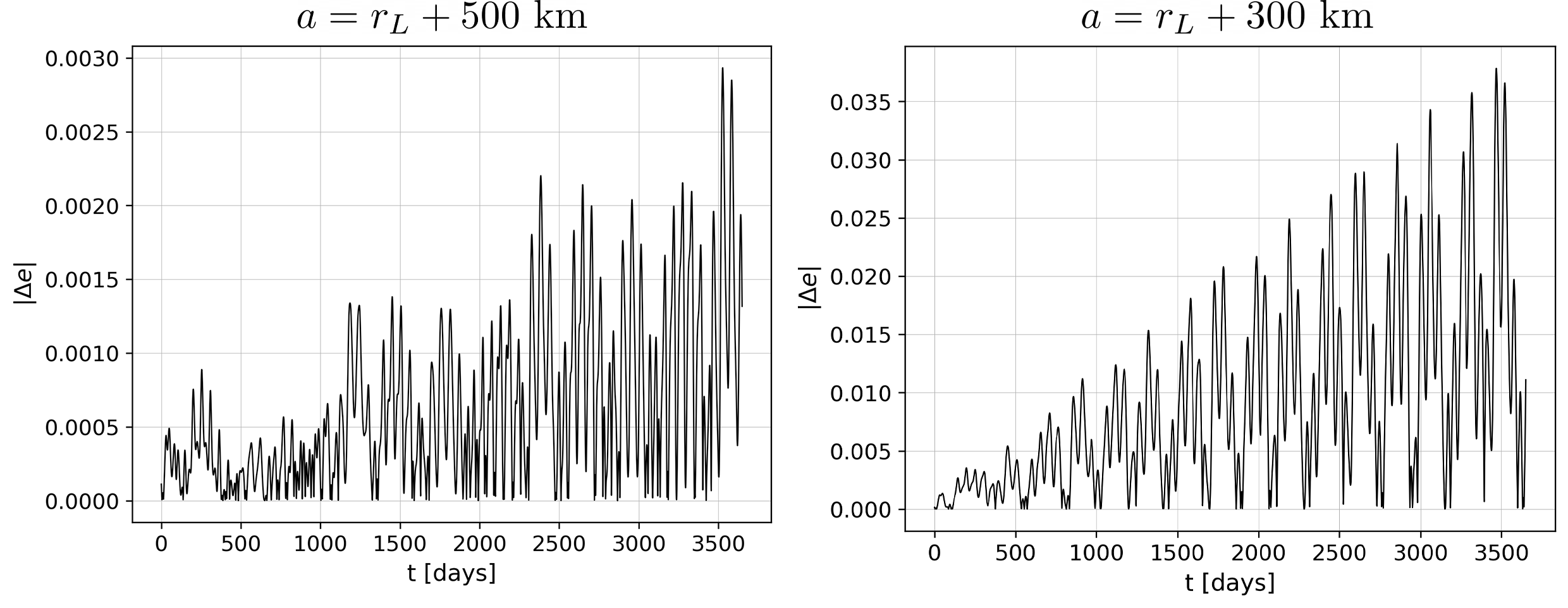}
\caption{\small The error in eccentricity $|\Delta\ue(t)|$ of the analytical versus numerical propagation of the mean elements as a function of time, shown in linear scale, for the trajectories with initial mean elements $\uell(0)=0$, $\ug(0)=-0.4$ rad, $\uh(0)=0.7$ rad, $\ue(0)=0.1$, $\ui(0)=15^\circ$, and (left) $\ua=500$~km, (right) $\ua=300$~km.
}
\label{fig:errorlinear}
\end{figure}

A quick visual inspection of Figs.~\ref{fig:errormapsecc} and ~\ref{fig:errormapsinc} shows that the level of the error of the analytical propagator is $\sim 10^{-4}$--$10^{-3}$ or smaller for most initial conditions at altitudes $500$ km$\leq\delta a\leq 2000$ km. As an example, Fig.~\ref{fig:orbitsgood} shows the comparison between the evolution of the mean eccentricity and mean inclination $(\ue(t),\ui(t))$ for the trajectories with initial conditions $\ue(0)=0.1$, $\ui(0)=15^\circ$ and semi-major axes $a[$km$]=R_{\Moon}+500$, $700$, $900$, $2000$. Besides quasi-periodic oscillations, with a period of order of several months, a small systematic trend in the growth of the error for these trajectories can be identified, which is better revealed plotting the errors in linear rather than logarithmic scale, as in Fig.~\ref{fig:errorlinear}. This error actually represents an error in the analytically computed secular frequencies $\nu_\gamma(a_p,e_p,i_p)$, $\nu_\theta(a_p,e_p,i_p)$, due, in turn, to the error in the transformation $\mathcal{X}_p$ used to compute the value of the proper elements $(e_p, i_p)$ starting from the initial values of the mean elements $\ue(0)$, $\ui(0)$. Through the last two of Eqs.~\eqref{propersol}, the error in the analytical values of the secular frequencies generates a gradual dephasing of the analytical solution from the numerical one. As a rule of thumb, for a propagation over a period of $~10$~yrs, the systematic error due to dephasing increases by about one order of magnitude with respect to the errors illustrated in the maps of Figs.~\ref{fig:errormapsecc} and~\ref{fig:errormapsinc}. 

\begin{figure}[!htp]
\centering
\includegraphics[scale=0.6]{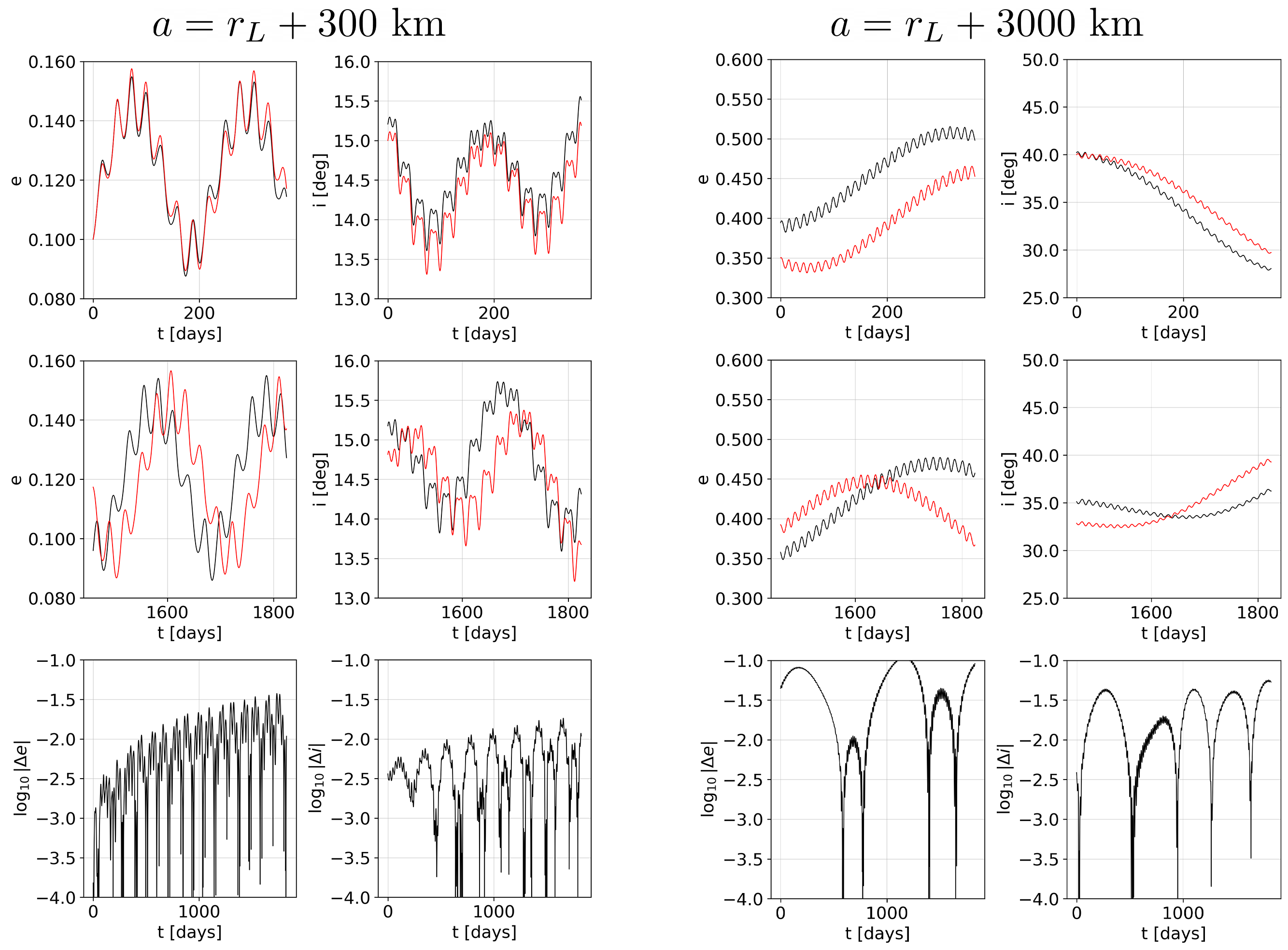}
\caption{\small   
Analytical (black line) versus numerical (red line) propagation of the mean elements $(\ue(t),\ui(t))$ for two orbits with groups of panels similar as in Fig.~\ref{fig:orbitsgood} and initial conditions $\uell(0)=0$, $\ug(0)=-0.4$~rad, $\uh(0)=0.7$~rad and (left) $\ua=R_{\Moon}+300$~km, $\ue(0)=0.1$, $\ui(0)=15^\circ$, (right) $\ua=R_{\Moon}+3000$~km, $\ue(0)=0.35$, $\ui(0)=40^\circ$.
}
\label{fig:orbitsbad}
\end{figure}

Returning to the latter maps, we notice however, that the errors increase significantly in particular domains of initial conditions. We distinguish two cases: \\
\\
\noindent
- the short term ($\sim 1$ year) propagation is precise, but the long term ($\sim 10$ years) propagation leads to a cumulative error of order $\sim 10^{-1}$. An example is the trajectory whose evolution of mean elements is shown in the left group of panels in Fig.~\ref{fig:orbitsbad}. The dephasing is now visually noticeable after 5 years of propagation and leads to an error growth as shown in the second panel of Fig.~\ref{fig:errorlinear} (or, equivalently, in the last-left panel of Fig.~\ref{fig:orbitsbad} in logarithmic scale). With reference to Fig.~\ref{fig:errormapsecc} and \ref{fig:errormapsinc}, this type of error mostly concerns the orbits in the domain $\ui(0)<20^\circ$ in the first panel of these figures, i.e., for the low altitude orbits $\ua=R_{\Moon}+300$~km. The main source of error is this case stems from the not so good convergence of the secular normal form, due to the fact that the orbits undergo oscillations in eccentricity that lead to a maximum eccentricity $\ue_{max}$ implying a minimum pericentric distance $r_{p,min}=\ua(1-\ue_{max})$ close to the Moon's surface, hence outside the zone of essentially secular orbits (see subsection \ref{subsec:potential}). For example, for the orbit of Fig~\ref{fig:orbitsbad}, left, we have $\ue_{max}\simeq 0.15$, implying $r_{p,min}\simeq R_{\Moon}$.\\
\\
- The analytical propagation contains a $\sim 10^{-1}$ error already from the beginning, and therefore cannot be considered valid. With reference to Fig.~\ref{fig:errormapsecc} and \ref{fig:errormapsinc}, this refers to the orbits with initial conditions in narrow yellow strips observed in all panels, together with the orbits seen in the top right part of the sixth panel (for $\ua=R_{\Moon}+3000$~km.) Example of the latter is the orbit in the right group of panels of Fig.~\ref{fig:orbitsbad}. Note that in this case the error is introduced already at the beginning of the process of analytical propagation. In particular, at time $t=0$ the analytical solution in mean elements is given by the transformation 
\begin{equation}\label{xanal0}
x_{analytical}(0)=\mathcal{X}_p^{-1}\left(\mathcal{X}_p(x(0))\right)\, .
\end{equation}
In theory, this is an identity transformation. In practice, however, there is an error due to the imprecise computation of the transformation $\mathcal{X}_{p}$ and of its inverse, which are given as truncated Lie series (Eqs.~\eqref{liechip} and \eqref{liechipinv}). In the case of the right panels of Fig.~\ref{fig:orbitsbad}, the error $x_{analytical}(0)-x(0)$ yields an initial difference $\Delta\ue(0)\simeq 5\times 10^{-2}$ which then propagates at all later times.

It is easy to verify that the most important source of the latter error is not the series truncation (which actually applies to all analytically computed trajectories), but the presence, in the transformation formulas, of \textit{small divisors} due to \textit{secular resonances}. We now focus on an analysis of these resonances.

\subsection{Secular resonances and small divisors}
\label{subsec:resonances}
\begin{figure}
\centering
\includegraphics[scale=0.65]{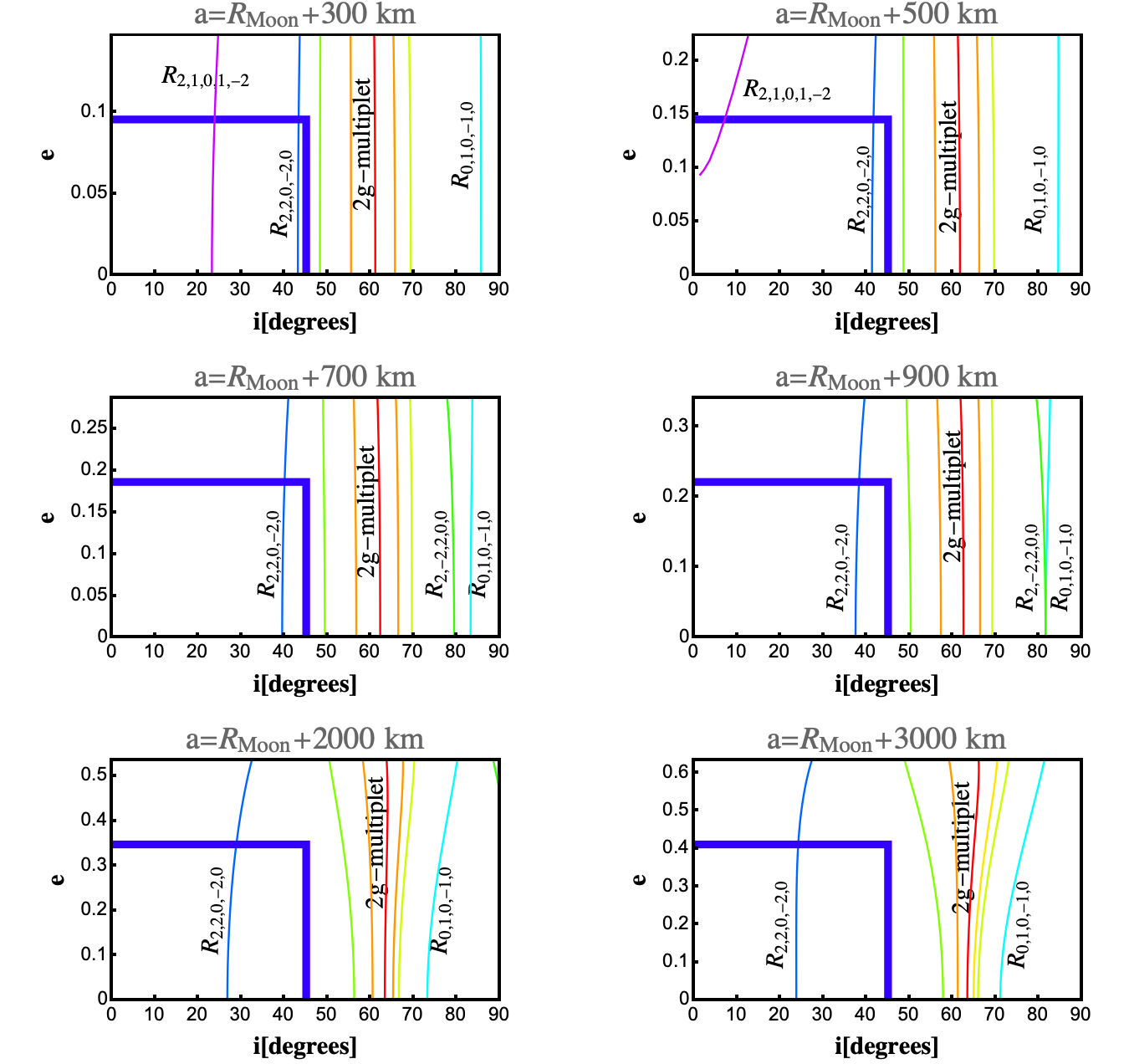}
\caption{\small 
The curves $e_p(i_p)$ of all the dynamically important resonances for prograde orbits ($i<90^\circ$) and for the set of values of the semi-major axis $a_p=\ua  = R_{\Moon}+\delta a$ with $\delta a= 300$, $500$, $700$, $900$, $2000$, $3000$~km. The limits in eccentricity in each plot is $0\leq e_p\leq \delta a/\ua$, same as in Figure 5 of \cite{legeft2024}. The parallelogram with thick blue border indicates the limits of the corresponding panels in Figs.~\ref{fig:errormapsecc} and \ref{fig:errormapsinc} of the present paper. The parallelogram is crossed by the resonances $\Rscr_{2,1,0,1,-2}$ (resonant angle: $2g+\widetilde{\Omega}+h_{\Moon}-2\ell_{\Sun}$, purple), for $a_p=R_{\Moon}+300$~km and $a_p=R_{\Moon}+500$~km, and $\Rscr_{2,2,-2,0,0}$ (resonant angle: $2g+2\widetilde{\Omega}-2\varpi_{\Moon}$, blue), for the entire set of values of $a_p$. The corresponding curves compare well with the centers of the narrow yellow strips found in the panels of equal value of the semi-major axis shown in Figs.~\ref{fig:errormapsecc} and \ref{fig:errormapsinc}. Outside the parallelogram, at all values of the semi-major axis, we have the bundle of resonances collectively referred to as `2g-multiplet',  with the `2g-resonance' ($\Rscr_{2,0,0,0,0}$, red) in the center, and the resonances $2g+(\widetilde{\Omega}-h_{\Moon})$ (first left), $2g-(\widetilde{\Omega}-h_{\Moon})$ (first right), $2g+2(\widetilde{\Omega}-h_{\Moon})$ (second left), $2g-2(\widetilde{\Omega}-h_{\Moon})$ (second right). Finally, in the domain of polar orbits we have the secular resonance $\widetilde{\Omega}-h_{\Moon}$ ($\Rscr_{0,1,0,-1,0}$), which appears at inclinations near $90^\circ$ at low altitudes and approaches to the 2g-multiplet as the altitude increases, as well as the transient resonance $2g-2\widetilde{\Omega}+2\varpi_{\Moon}$ (i.e. $\Rscr_{2,-2,2,0,0}$, which appears at the altitude $~700$km and disappears at the altitude $\sim 2000$km.)
}
\label{fig:resonances}
\end{figure}
\begin{figure}[htp]
\centering
\includegraphics[scale=0.5]{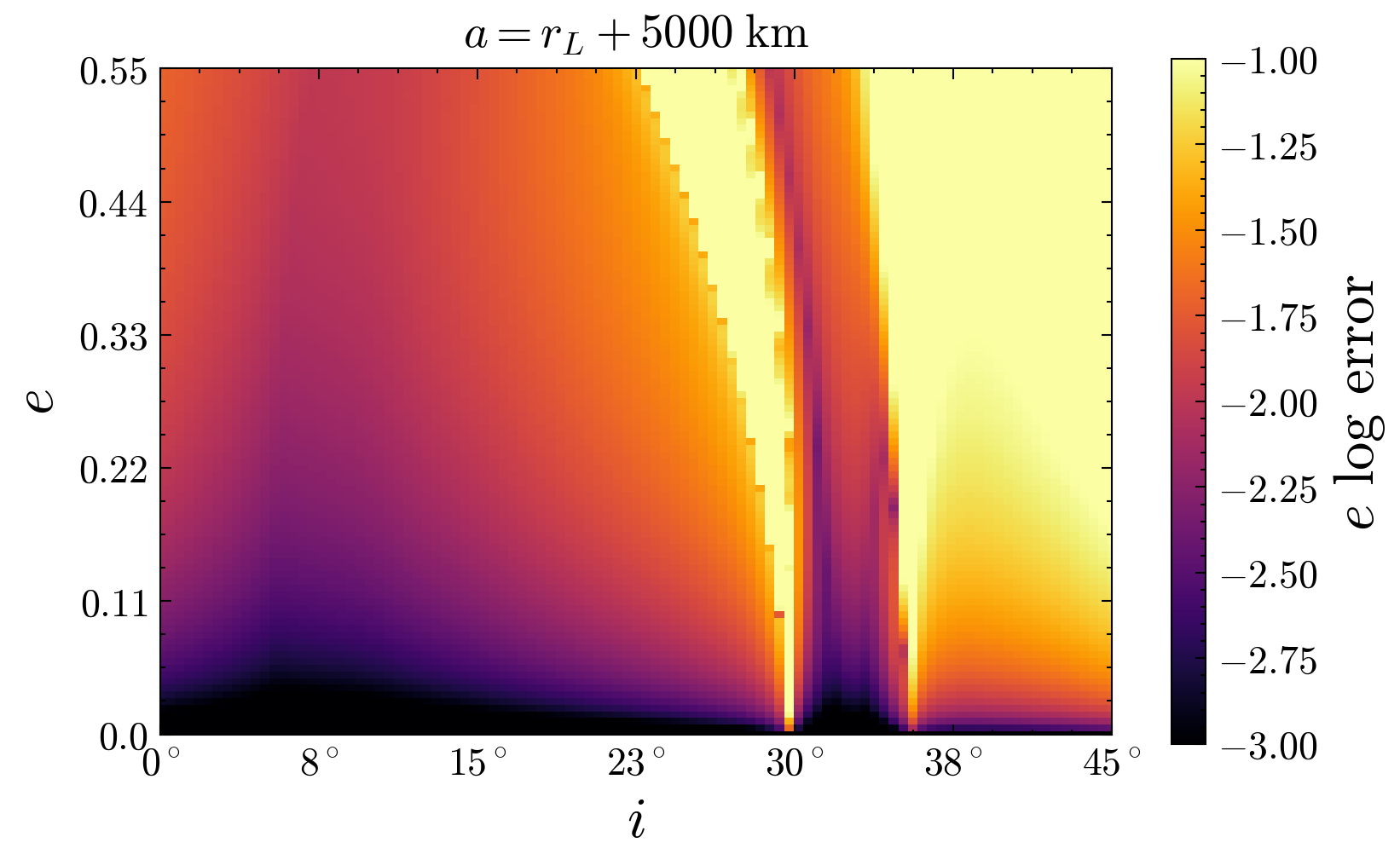}
\includegraphics[scale=0.5]{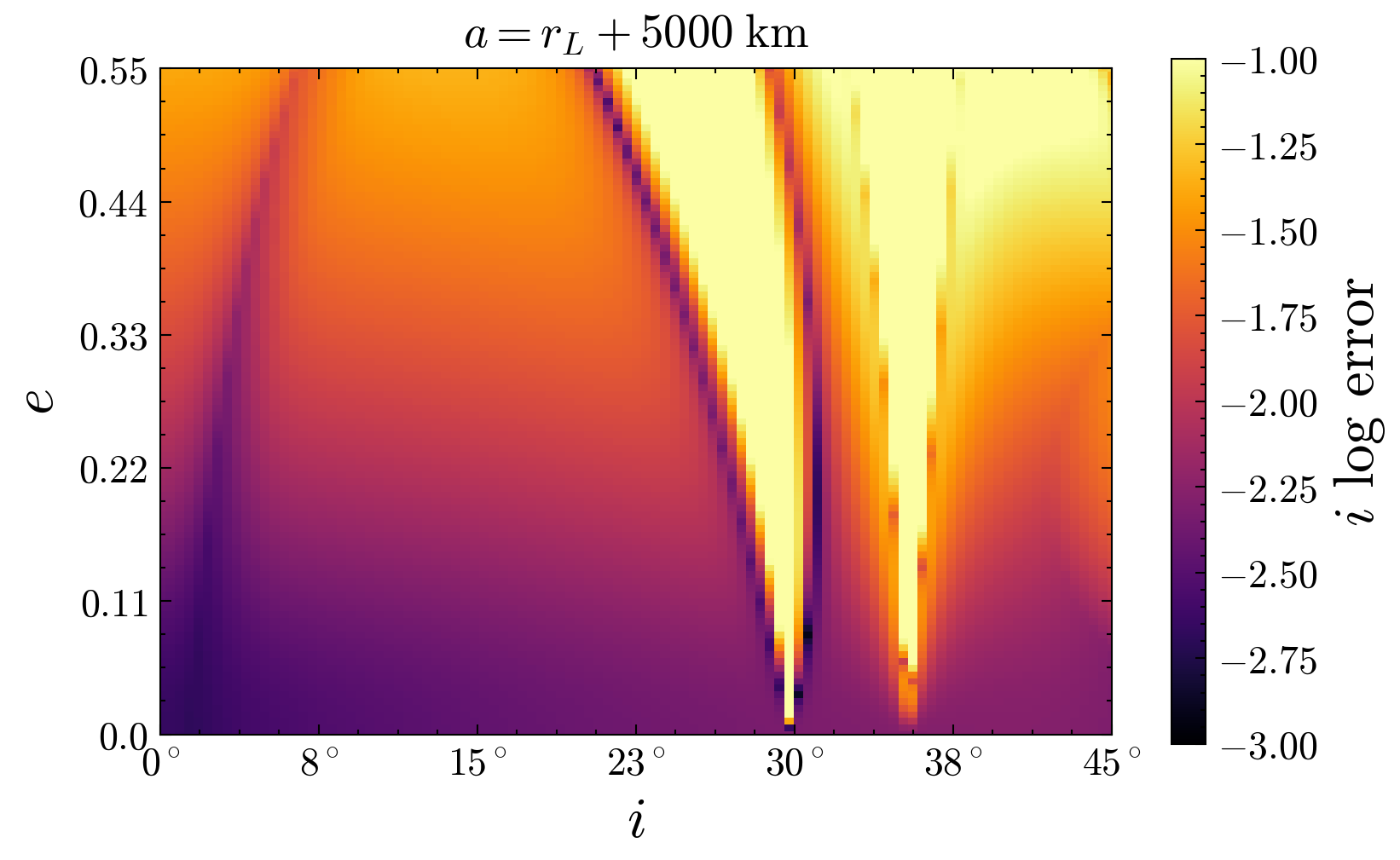}
\caption{
\small Maps of the errors $\log_{10}(|\Delta\ue(t)|)$ (left) and $\log_{10}(|\Delta\ui(t)|)$ (right), as in Figs.~\ref{fig:errormapsecc} and \ref{fig:errormapsinc} but at the semi-major axis $\ua=R_{\Moon}+5000$~km. 
}
\label{fig:errormapsa5000km}
\end{figure}
\begin{figure}[!h]
\centering
\includegraphics[scale=0.6]{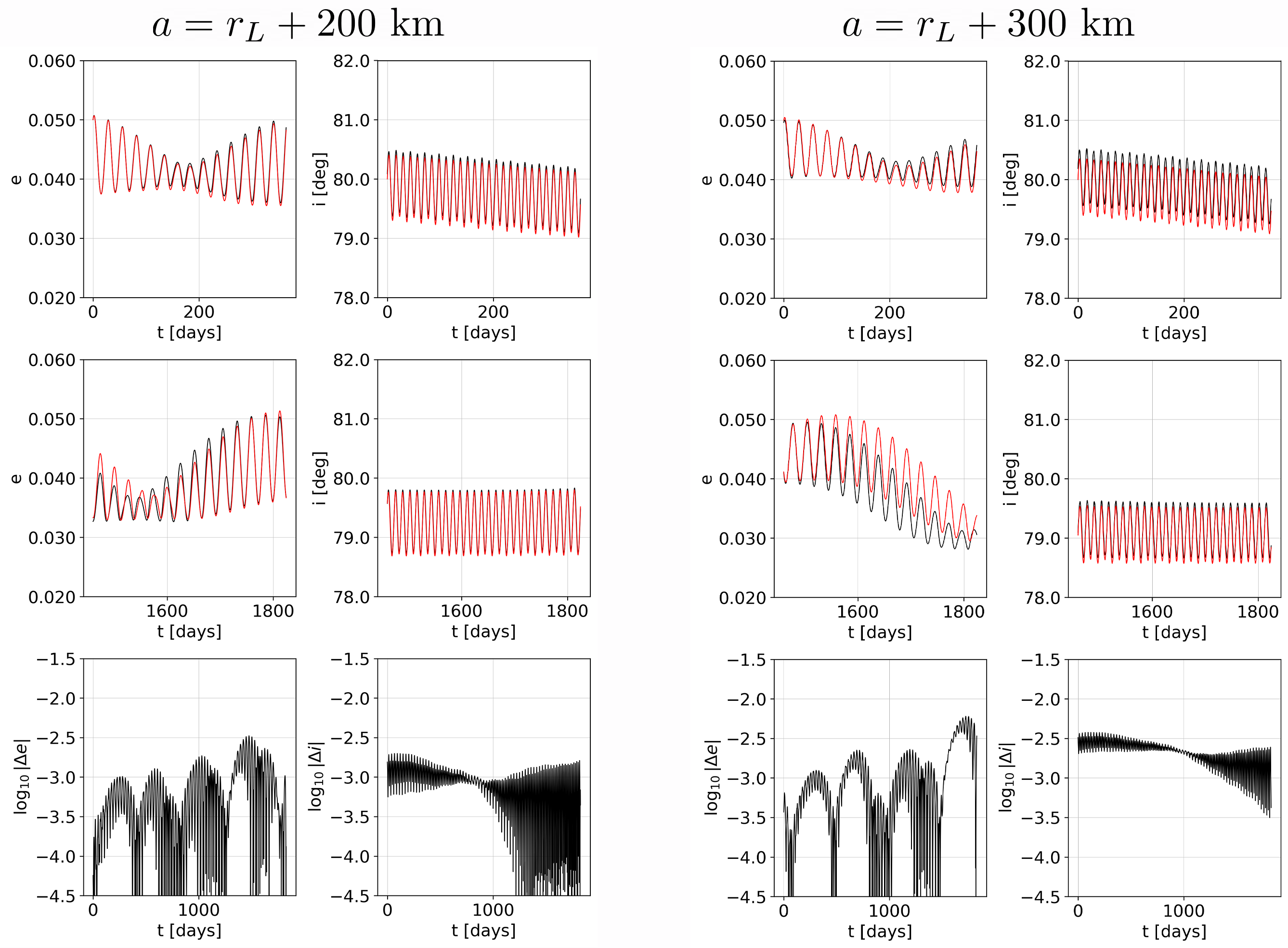}
\caption{
\small Two nearly polar orbits with initial conditions $\uell(0)=0$, $\ug(0)=-0.4$ rad, $\uh(0)=0.7$ rad, $\ue(0)=0.05$, $\ui(0)=80^\circ$, and (left) $\delta a=200$~km, (right) $\delta a=300$~km. The initial conditions of the orbit in left are sufficiently far from the $\widetilde{\Omega}-h_{\Moon}$ resonance, while the orbit in right is slightly affected by the latter resonance. Owing to their low eccentricity, both orbits are protected from close approaches to the Moon's surface.}
\label{fig:orbitspolar}
\end{figure}

Let $\ua$ be a fixed (and constant in time) choice of value of the mean semi-major axis. Setting the proper semi-major axis equal to $a_p=\ua$, a point in the plane of proper elements $(i_p,e_p)$ is said to satisfy the exact \textit{secular resonance} $\Rscr_{m_1,m_2,m_3,m_4,m_5}$ if the following commensurability holds between the five secular frequencies of the problem:
\begin{equation}\label{secres}
\begin{aligned}
\Rscr_{m_1,m_2,m_3,m_4,m_5}:~&m_1\dot{g}(\ua,e_p,i_p)+
m_2(\dot{h}(\ua,e_p,i_p)+\nu_1)+
m_3\nu_2+
m_4\nu_3+
m_5\nu_4 =\\
-&m_1\nu_\gamma(\ua,e_p,i_p)
+(m_1-m_2)\nu_\theta(\ua,e_p,i_p)+
m_3\nu_2+
m_4\nu_3+
m_5\nu_4=0~~.
\end{aligned}
\end{equation}
A secular resonance $\Rscr_{m_1,m_2,m_3,m_4,m_5}$ is called \textit{dynamically important} if:
\begin{itemize}
\item
in the trigonometric part of the secular Hamiltonian~\eqref{hamsec1mod} there exists a non-null harmonic, or multiple thereof, with $k_1'=-m_1$, $k_2'=(m_1-m_2)$, $m=m_3$, $l=m_4$, $s=m_5$,
\item
the curve $e_p(i_p)$ for which the condition (\ref{secres}) is satisfied crosses the domain $0\leq e_p\leq \delta a/\ua$ with $0\leq i_p\leq 90^\circ$ for prograde orbits, or $90^\circ<i_p<180^\circ$ for retrograde orbits.
\end{itemize}  

Figure \ref{fig:resonances} shows the curves $e_p(i_p)$ of all the dynamically important secular resonances for prograde orbits and values of the semi-major axis $a_p=\ua=R_{\Moon}+\delta a\,,$ where $\delta a= $ $300$, $500$, $700$, $900$, $2000$, $3000$~km, same as those of the consecutive panels in Figs.~\ref{fig:errormapsecc} and ~\ref{fig:errormapsinc}. For reasons of comparison, the limits of the outer borders of all panels in Fig.~\ref{fig:resonances} are equal to those of the corresponding (in semi-major axis) panels in Figure 5 of \cite{legeft2024}. Also, the borders of the inner parallelogram defined in each panel by the thick-blue inserted lines coincide with the borders of the corresponding panel in Figs.~\ref{fig:errormapsecc} and \ref{fig:errormapsinc}.

Direct visual comparison allows now to see that the centers of all the double yellow strips observed in Figs.~\ref{fig:errormapsecc} and \ref{fig:errormapsinc} correspond to the resonant curves produced by only two secular resonances. Denoting by $\widetilde{\Omega}$ the slow angle $\widetilde{\Omega}=-\varphi_\theta=h+\lambda_{\Moon}$, we refer to these two resonances as the `$2g+\widetilde{\Omega}+h_{\Moon}-2\ell_{\Sun}$' resonance ($\Rscr_{2,1,0,1,-2}$, purple curve), and the `$2g+2\widetilde{\Omega}-2\varpi_{\Moon}$' resonance ($\Rscr_{2,2,-2,0,0}$, blue curve). For $a_p=\ua=R_{\Moon}+300$~km, the first of these two resonances appears as a nearly vertical strip in Figs.~\ref{fig:errormapsecc} and \ref{fig:errormapsinc} at the inclination $\approx 24^\circ$. At higher altitudes, the resonance moves towards the upper left part of the permissible $(\ui,\ue)$ plane. Hence at the semi-major axis $a_p=\ua=R_{\Moon}+500~$km the resonance yields a double arc in the upper-left corner of the plot, and at higher altitude the curve of the resonance goes outside the considered domain. On the other hand, the $2g+2\widetilde{\Omega}-2\varpi_{\Moon}$ resonance is present in all panels of Figs.~\ref{fig:errormapsecc} and \ref{fig:errormapsinc} as a nearly vertical double strip which moves to lower levels of inclination at higher altitudes. The fact that the error plots yield double strips instead of single curves (as those of Fig.~\ref{fig:resonances}) is explained by the general theory of secular resonances for artificial satellites (see, for example,~\cite{legeft2023}), and it is a consequence of the fact that these strips yield the thin separatrix-like chaotic layers of resonances of the `second fundamental model', as explained in~\cite{brei2001}. Note also that the coincidence between the centers of the double strips in Figs.~\ref{fig:errormapsecc} and \ref{fig:errormapsinc} and the resonant curves in Fig.~\ref{fig:resonances} is only approximate, as, contrary to the \textit{equality} $\ua=a_p$, the transformation $\mathcal{X}_p$ is only a \textit{near-identity} for the mean and proper elements $\ue\approx e_p$, $\ui\approx i_p$. In fact, the exact relation depends also on the secular angles $(\ug,\uh)$ and on the epoch, i.e., the angles $\tilde{\varphi}_1$ and $\varphi_j$, $j=2,3,4$. Finally, the error in the analytical propagation of the trajectories with initial conditions within the above resonant strips is due to the fact that, through the dependence of the divisors of Eq.~\eqref{chip} (explicitly reported in Eq.~\eqref{divisors}) on the eccentricity and inclination, the transformations $\mathcal{X}_p$ and $\mathcal{X}_p^{-1}$ of the analytical theory yield terms with divisors very small or exactly equal to zero. Such cases are singular and are not covered by the analytical propagator. 

Outside, now, the inner parallelogram, we have a second group of five resonances which, in all panels of Fig.~\ref{fig:resonances}, form a nearly vertical bundle around the `2g-' or `critical' resonance ($\Rscr_{2,0,0,0,0}$). We defer to a future paper a detailed discussion of the resonances adjacent to the critical one, noting only that in most cases the whole bundle is within the domain of re-entry orbits found in Figure 5 of \cite{legeft2024}. Such orbits, in turn, are due entirely to the structure of the separatrices of the critical resonance, as explained in \cite{legeft2024}. Furthermore, the interaction between the resonances of the bundle and the $2g+2\widetilde{\Omega}-2\varpi_{\Moon}$ appears to create an extended resonance overlap domain in Figs.~\ref{fig:errormapsecc} and \ref{fig:errormapsinc} for $a_p=\ua>2000~$km, which covers all orbits with inclinations in the approximate range $30^\circ\sim 70^\circ$, as also seen in Fig.~\ref{fig:errormapsa5000km}.

Finally, at low altitudes and inclinations beyond the domain of the 2g-multiplet, we can still have regular nearly polar orbits, which are reproduced by the analytical propagator with sufficient accuracy.  As seen in Fig.~\ref{fig:orbitspolar}, such orbits must have initial conditions not very close to the $\widetilde{\Omega}-h_{\Moon}$ resonance ($\Rscr_{0,1,0,-1,0}$), which is a resonance generated at low altitudes at inclinations near $90^\circ$, and moves towards the 2g-multiplet zone at higher altitudes (see Fig.~\ref{fig:resonances}), as well as the transient resonance $2g-2\widetilde{\Omega}+2\varpi_{\Moon}$ ($\Rscr_{2,-2,2,0,0}$) which appears at the altitude $700$~km, merges with the domain of re-entry orbits at the altitude $\sim 1500$~km, and disappears at the altitude $\sim 2000$~km.

\subsection{Accuracy tests in comparison with a Cartesian propagator}

We finally test the accuracy of the whole analytical procedure described in subsection \ref{subsec:symbolic}, by which the osculating elements of a trajectory at time $0$ are mapped analytically to the osculating elements at time $t$. Formally, let $\Pscr_{t_0,t}$ by the analytical mapping which propagates all proper variables $\mathbf{x}'$ from their values at time $t_0$ to the ones at time $t$ through Eqs.~\eqref{propersol} and \eqref{phases}:
\begin{equation}\label{pmap}
\mathbf{x}'(t)=\Pscr_{t_0,t}(\mathbf{x}'(t_0))~~.
\end{equation}
The whole procedure of analytical propagation can then be resumed as follows: define the vector of osculating variables as 
$$
\mathbf{z}=(a,e,i,\ell,g,h,\varphi_1,\varphi_2,\varphi_3,\varphi_4)~~.
$$
The analytical propagation corresponds to the chain composition of the operations $\mathbf{x}(t_0)=\mathcal{X}(\mathbf{z}(t_0))$, $\mathbf{x'}(t_0)=\mathcal{X}_p(\mathbf{x}(t_0))$, $\mathbf{x'}(t)=\Pscr_{t_0,t}(\mathbf{x}'(t_0))$, $\mathbf{x}(t)=\mathcal{X}_p^{-1}(\mathbf{x'}(t))$ and $\mathbf{z}(t)=\mathcal{X}^{-1}(\mathbf{x}(t))$, or:
\begin{equation}\label{analmap}
\Ascr_{t_0,t}:~~\mathbf{z}(t_0)\rightarrow\mathbf{z}(t)~~\mbox{where}~~
\mathbf{z}(t)=
\left(\mathcal{X}^{-1}\circ\mathcal{X}_p^{-1}\circ\Pscr_{t_0,t}
\circ\mathcal{X}_p\circ\mathcal{X}\right)
(\mathbf{z}(t_0))~~.
\end{equation}

\begin{figure}[!htp]
\centering
\includegraphics[scale=0.525]{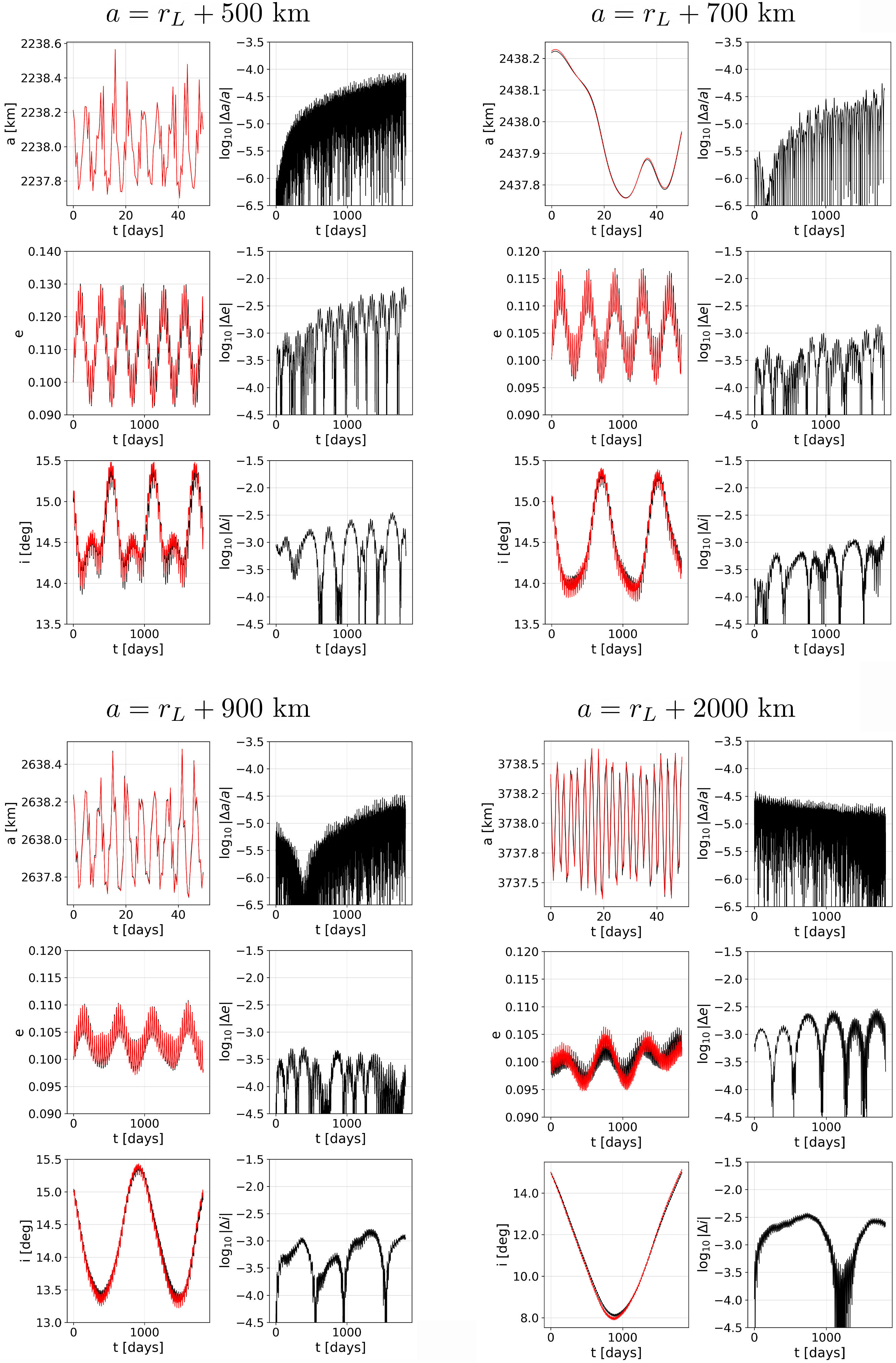}
\caption{\small   
Analytical (black line) versus numerical (with a cartesian propagator, red line) propagation of the osculating elements $(a(t),e(t),i(t))$ for four orbits with initial conditions similar to those of Fig.~\ref{fig:orbitsgood}, but now in osculating elements $\ell(0)=0$, $g(0)=-0.4$~rad, $h(0)=0.7$~rad, $e(0)=0.1$, $i(0)=15^\circ$,  and semi-major axis corresponding to the group of panels $a(0)=R_{\Moon}+500$~km (top left), $a(0)=R_{\Moon}+700$~km (top right), $a(0)=R_{\Moon}+900$~km (bottom left), $a(0)=R_{\Moon}+2000$~km (bottom right). In each group of panels, the first row shows the comparison analytical versus numerical for the osculating value of the semi-major axis $a(t)$ (left panel), and the corresponding relative error in $\log_{10}$ scale (right panel), while the second and third rows show the same comparisons for the orbit's evolution of the osculating value of the eccentricity $e(t)$ and inclination $i(t)$ respectively.  
}
\label{fig:orbitsgood_cart}
\end{figure}

As shown in \cite{eftetal2023}, the errors in the transformations $\mathcal{X}$, $\mathcal{X}^{-1}$ are very small, typically of order (in the relative error) $10^{-6}\sim 10^{-5}$. Instead, as shown in the previous subsection, the errors in the transformations $\mathcal{X}_p$, $\mathcal{X}^{-1}_p$ are in general larger, due to the presence of small divisors in the corresponding Lie series. Hence, far from secular resonances, the relative errors are of order $10^{-4}\sim 10^{-3}$, while for initial conditions close to secular resonances they increase rapidly to the level of $10^{-1}$, rendering the analytical solutions non applicable. Note, however, that these errors do not affect the evolution of the semi-major axis, which is a constant of motion in mean as well as in proper elements. In synthesis, the error of the analytical propagation of the osculating elements (mapping $\Ascr_{t_0,t}$, described in Eq.~\eqref{analmap}) is dominated by the error in the transformation $\mathcal{X}_p$ for all the elements except for the osculating value of the semi-major axis, for which the error is generated only by the transformation $\mathcal{X}$. 

\begin{figure}[!h]
\centering
\includegraphics[scale=0.6]{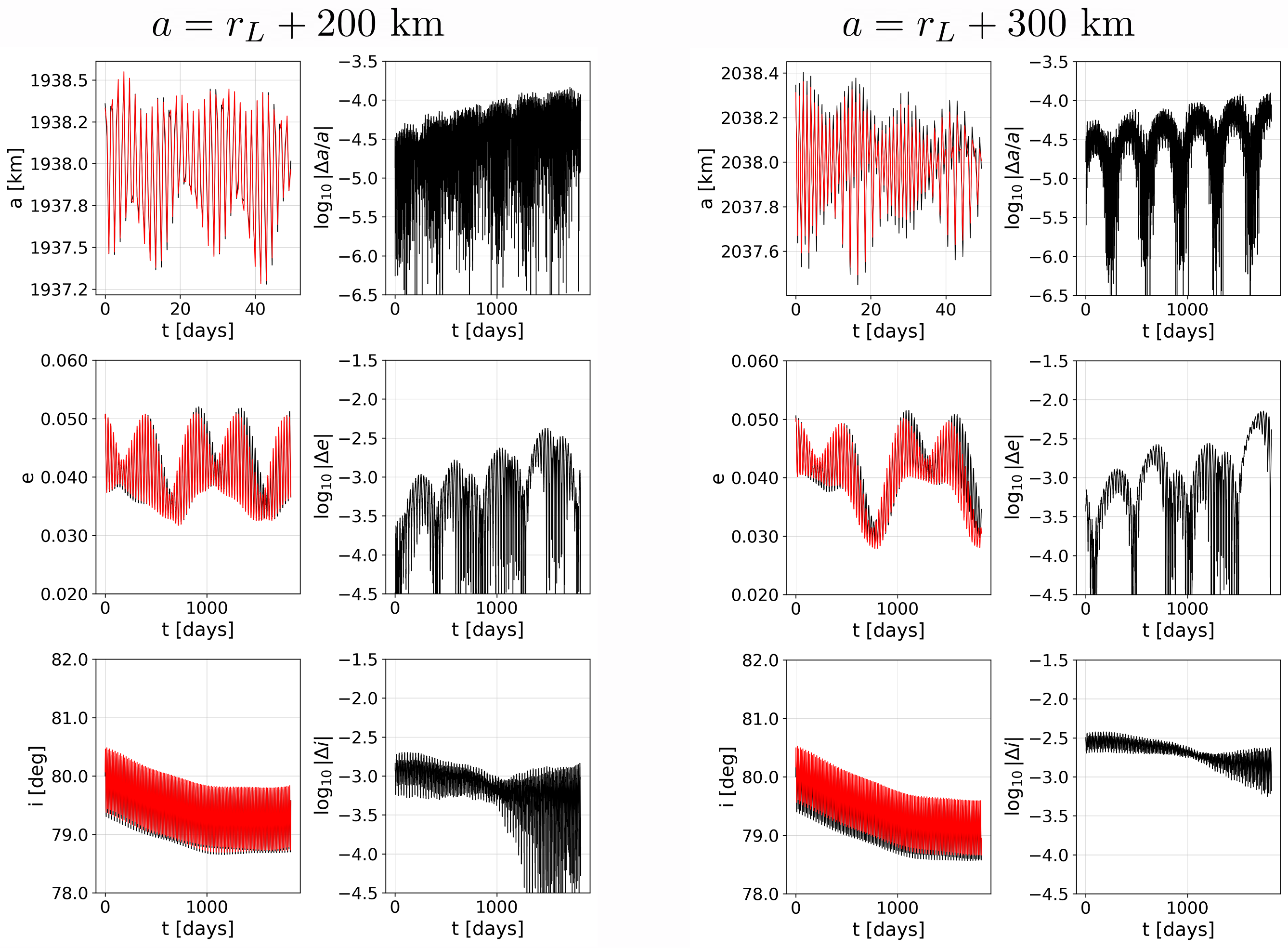}
\caption{\small   
Comparison of the analytical versus numerical propagation of the osculating elements $(a(t),e(t),i(t))$ (same as in Fig.~\ref{fig:orbitsgood_cart}, but for two polar orbits with initial conditions similar to those of Fig.~\ref{fig:orbitspolar}, but now in osculating elements, namely $\ell(0)=0$, $g(0)=-0.4$~rad, $h(0)=0.7$~rad, $e(0)=0.05$, $i(0)=80^\circ$,  and semi-major axis $a(0)=R_{\Moon}+200$~km (left) or $a(0)=R_{\Moon}+300$~km (right)).   
}
\label{fig:orbitspolar_cart}
\end{figure}

This analysis is verified through a direct comparison of orbits whose osculating elements are propagated fully analytically, through the mapping $\Ascr_{t_0,t}$, or fully numerically, by a numerical integration of the equations of motion using Cartesian coordinates and velocities for the satellite's orbit. These numerical integrations have been made using the \textit{heyoka} python library~\cite{bisizz2021}. Figure~\ref{fig:orbitsgood_cart} yields the relevant information. The figure shows a comparison of the fully analytical with the fully numerical orbit as regards the evolution of the osculating elements $a(t)$, $e(t)$, $i(t)$, in the case of four trajectories chosen with initial conditions similar to those of Fig.~\ref{fig:orbitsgood}, but now taken in osculating rather than mean Keplerian elements. We note that, now, the osculating semi-major axis is no longer constant, but undergoes small amplitude oscillations which are dominated by the orbit's shortest period (mean motion frequency $(\cgrav M_{\Moon}/a^3)^{1/2}$), as well as by the Moon's (semi-secular) monthly period. Since for the semi-major axis the only error in the analytical solution is due to the transformations $\mathcal{X}$, $\mathcal{X}^{-1}$, we obtain very small relative errors which grow in time from an initial level of $10^{-6}$ to $10^{-4}$ after five years of propagation. On the contrary, as regards the error in the comparison of the analytical versus numerical orbit for the osculating elements $e(t)$ and $i(t)$, we obtain relative errors essentially at the same level as those observed in Fig.~\ref{fig:orbitsgood}, i.e., the error here is dominated by the transformations $\mathcal{X}_p$, $\mathcal{X}^{-1}_p$ of the secular theory. A similar picture is recovered in the domain of polar orbits, as concluded from the comparison of Fig.~\ref{fig:orbitspolar_cart} with Fig.~\ref{fig:orbitspolar}.  

\section{Conclusions}
\label{sec:conclusion}
In this paper we present the theory, as well as an open source library, of a fully-analytical long-term (i.e. in timescale of decades) propagator of lunar satellite orbits. Our propagator is based on a combination of a previously developed (in \cite{eftetal2023}) semi-analytical theory, as well as a secular theory developed in section \ref{sec:analytical} of the present paper. The former theory introduces a near-to-identity transformation from osculating to mean elements (transformation $\mathcal{X}$, Eq.~\eqref{tramean}) which allows to reproduce analytically all short-period variations of a satellite orbit's Keplerian elements. The theory developed in the present paper, instead, is based on a near-to-identity transformation from mean to proper elements (transformation $\mathcal{X}_p$, Eq.~\eqref{liechip}), which yields integrable equations of motion in the proper elements. Our main validation steps for this propagator and thereby drawn conclusions are the following:
\begin{enumerate}
\item
We use as a point of departure a lunar + Earth gravity model which essentially consists of the same set of harmonics of the lunar gravity model as proposed in \cite{legeft2024}, and the quadrupolar terms of the Earth's gravity computed in the Moon's co-rotating PALRF frame. However, we modify the model used for the Earth's lunicentric position vector (see subsection \ref{subsec:potential}) with respect to the one used in \cite{legeft2024}, including now the harmonics related to this vector which involve the solar year frequency as well as the frequencies of precession of the Moon's pericenter and line of nodes in the ecliptic plane. As argued in subsection {\ref{subsec:resonances}}, introducing these frequencies is crucial in understanding the long-term orbital dynamics, since it leads to the appearance of secular resonances which greatly alter the nature of the orbits, in particular at inclinations beyond $45^\circ$. As a general guide, for prograde orbits our analytical propagator is accurate (at a relative error level $10^{-4}\sim 10^{-3}$) at all altitudes $\delta a=a-R_{\Moon}$ in the range $300~\mbox{km}\leq \delta a\leq 3000~\mbox{km}$, for almost all initial conditions in the square $0\leq i\leq 45^\circ$, $0\leq e\leq 0.7(a-R_{\Moon})/a$, except for thin narrow zones around the secular resonances $2g+\widetilde{\Omega}+h_{\Moon}-2\ell_{\Sun}$ and $2g+2\widetilde{\Omega}-2\varpi_{\Moon}$ (see subsection \ref{subsec:resonances}). 

\item
For initial conditions outside the above indicated domain, the nature of the orbits as well as the limits of applicability of our propagator are determined by the effect of three independent approximations made in our model. Thus: i) for low altitude orbits ($\delta a\lesssim 300~\mbox{km}$), secular variations in the eccentricity lead to orbits with pericenters close to the lunar surface. In this case, our transformation $\mathcal{X}_p$ has poor convergence, and we obtain analytical solutions with a cumulative error growing to the level of $\sim 10^{-1}$ after about one year of propagation. Also, the truncation of the lunar gravity model to the set of harmonics as in Eq.~\eqref{ssmset} becomes problematic, since lunar mascons render many more harmonics relevant to the problem. However, the propagator remains accurate at altitudes $100~\mbox{km}\leq\delta a\leq 300~\mbox{km}$ for orbits which do not undergo large variations of the eccentricity, as, for example, nearly polar orbits, provided that these orbits are taken with initial conditions far from the $\widetilde{\Omega}-h_{\Moon}$ resonance. ii) At high altitudes $\delta a>3000~\mbox{km}$ the extent of the resonant + resonance overlap domains of some secular resonances becomes important, and leads to substantial errors in the analytical estimates of the initial values of the proper elements before propagation. Note also that at such amplitudes the octupole and sextipole terms of the Earth's tide (P3 and P4) become relevant, while, at still higher altitudes, even solar gravity and radiation pressure may play a role. Nonetheless, the analytical method discussed in section \ref{sec:analytical} of the present paper is straightforward to apply including any additional gravity terms (lunar or third body) to the analytical propagator. iii) Beyond the inclination of $45^\circ$, the nature of dynamics is essentially determined by the extent of the separatrices of the 2g-resonance, as well as a set of additional resonances which are close to the 2g-resonance and form a `multiplet'. As a coarse division, we distinguish orbits which collide with the Moon's surface and others which do not. The corresponding limits are discussed in \cite{legeft2024}, and visualized in their Figure 5. For non-colliding orbits, our propagator is precise, provided that the initial conditions are taken far from the resonances discussed in Fig.~\ref{fig:resonances} of subsection \ref{subsec:resonances} above. 
\item
We make error tests aiming to quantify the level of error of the analytical propagator, first in the context exclusively of the secular variables (mean or proper) of the theory developed in the present paper, and then in the passage from mean to osculating elements by the semi-analytical theory of \cite{eftetal2023}. Due to the presence of small divisors in the transformation series of the present secular theory, our transformation $\mathcal{X}_p$ yields the dominant error of the analytical propagator as regards the evolution of the secular quantities (eccentricity and inclination vectors or, equivalently, the equinoctial elements $h_{eq},k_{eq},p_{eq},q_{eq}$ discussed in subsection \ref{subsec:symbolic}). On the other hand, our secular theory does not affect the evolution of the osculating value of the semi-major (whose either mean or proper value is constant). Hence, only the transformation $\mathcal{X}$ of the semi-analytical theory affects the relative error in the analytical propagation of the osculating value of the semi-major access, which, consequently, turns to be smaller than the error in the secular variables by about two orders of magnitude.

\end{enumerate}

\bibliographystyle{abbrvnat}
\bibliography{bibliography}

\end{document}